\documentclass[amsmath,amssymb]{revtex4}

\usepackage{graphicx}
\usepackage{dcolumn}
\usepackage{bm}

\newcommand{\bs}[1]{\mbox{\boldmath $#1$}}
\newcommand{\bcal}[1]{\mbox{\boldmath $\cal #1$}}

\begin{document}


\title{
Explicit Symplectic Integrators of Molecular Dynamics \\
Algorithms for Rigid-Body Molecules \\
in the Canonical, Isothermal-Isobaric, and Related Ensembles 
}

\author{Hisashi Okumura}
\email{hokumura@tb.phys.nagoya-u.ac.jp}
\author{Satoru G. Itoh}
\email{itoh@tb.phys.nagoya-u.ac.jp}
\author{Yuko Okamoto}
\email{okamoto@phys.nagoya-u.ac.jp}
\affiliation{%
Department of Physics \\
School of Science \\ 
Nagoya University \\
Furo-cho, Chikusa-ku \\
Nagoya, Aichi 464-8602, Japan}%



\begin{abstract}
We propose explicit symplectic integrators of 
molecular dynamics (MD) algorithms 
for rigid-body molecules in the canonical and isothermal-isobaric ensembles. 
We also present a symplectic algorithm in the constant normal pressure 
and lateral surface area ensemble 
and that combined with the Parrinello-Rahman algorithm. 
Employing the symplectic integrators for MD algorithms, 
there is a conserved quantity 
which is close to Hamiltonian. 
Therefore, we can perform a MD simulation more stably 
than by conventional nonsymplectic algorithms.
We applied this algorithm to a TIP3P pure water system at 300~K and 
compared the time evolution of the Hamiltonian 
with those by the nonsymplectic algorithms. 
We found that the Hamiltonian was conserved well 
by the symplectic algorithm even for a time step of 4~fs. 
This time step is longer than typical values of 0.5--2~fs 
which are used by the conventional nonsymplectic algorithms.
\end{abstract}

\maketitle

%
  \section{Introduction} \label{intro:sec}
%
There are two models for molecules in molecular dynamics (MD) simulations. 
One model is a rigid-body model and the other is a flexible model. 
Relative coordinates in a molecule are fixed in the rigid-body model, 
while they vary in the flexible model. 
Because degrees of freedom in the rigid-body model are fewer 
than in the flexible model, the simulational cost is less expensive. 
Several MD techniques have thus been proposed for rigid-body molecules. 

One possibility for the rigid-body modeling is to constrain 
a bond length and a bond angle among atoms in the molecules 
such as in the SHAKE algorithm \cite{ryck77}. 
Although it is easy to write a computational program for 
this constraint algorithm, 
it requires iteration procedures to fulfill the constraint. 
It means that one has to perform implicit time development. 

Another algorithm is a quaternion scheme 
which gives explicit time development. 
One integrator to carry out a quaternion MD simulation is 
the Gear's predictor-corrector algorithm \cite{gear71}. 
However, this algorithm is not a symplectic integrator \cite{yoshida90} 
nor time reversible.
It hardly reflects characteristics of Hamiltonian dynamics. 
Another algorithm for the quaternion MD was proposed by 
Matubayasi and Nakahara \cite{matu99}. 
Although this algorithm is not a symplectic integrator, 
it conserves volume in phase space and is time reversible. 
Miller $et$ $al$. recently proposed 
a symplectic quaternion algorithm \cite{mill02}. 
This algorithm also conserves volume in phase space and is time reversible. 
However, this symplectic quaternion algorithm has been proposed 
only in the microcanonical ensemble. 
There is no symplectic quaternion algorithm in the canonical ensemble and 
in the isothermal-isobaric ensemble. 

A representative MD algorithm to obtain the canonical ensemble 
is the Nos\'e thermostat \cite{nose84,nosejcp84}. 
Because the original Nos\'e Hamiltonian gives dynamics in virtual time, 
a symplectic canonical MD simulation can be carried out in virtual time. 
However, a symplectic MD simulation cannot be realized in real time. 
Nonsymplectic integrators such as Gear's predictor-corrector algorithm 
are often employed for real-time development for the Nos\'e thermostat. 
Hoover improved the Nos\'e thermostat to propose 
the Nos\'e-Hoover thermostat \cite{hoo85}. 
Because the Nos\'e-Hoover thermostat is not based on a Hamiltonian, 
there is no symplectic algorithm \cite{gold-symp} 
for the Nos\'e-Hoover thermostat. 
However, there exists an explicit time reversible integrator, 
although it does not conserve the volume in the phase space. 
This integrator was proposed by Martyna $et$ $al$. \cite{martyna96}. 
Bond $et$ $al$. then proposed 
a symplectic constant temperature algorithm in real time, 
which is refereed to as the Nos\'e-Poincar\'e thermostat \cite{bond99}. 
However, the original symplectic algorithm for 
the Nos\'e-Poincar\'e thermostat is an implicit integrator. 
Iterations are necessary for the thermostat. 
Nos\'e improved the original algorithm 
and proposed an explicit symplectic integrator 
for the Nos\'e-Poincar\'e thermostat \cite{nose01}.
Although this formalism may not be widely known, 
we found it very powerful and useful as was shown 
in Refs.~[\onlinecite{okokmd04,okokmd06}] and will be demonstrated below. 

An explicit MD algorithm closest to a symplectic algorithm 
for rigid-body molecules in the canonical ensemble 
proposed so far is a combined algorithm \cite{ike04} of 
the symplectic quaternion algorithm by Miller $et$ $al$. \cite{mill02} 
and time reversible algorithm 
for the Nos\'e-Hoover thermostat \cite{martyna96}. 
Because the Nos\'e-Hoover thermostat is a nonsymplectic algorithm, 
the whole algorithm is also nonsymplectic. 
Employing a symplectic MD algorithm, 
there is a conserved quantity which is close to Hamiltonian 
and the long-time deviation of the Hamiltonian is suppressed. 
Symplectic MD algorithms are thus getting popular recently. 

In this article, we propose an explicit symplectic MD algorithm 
for rigid-body molecules in the canonical ensemble 
by combining the quaternion algorithm by Miller $et$ $al$. \cite{mill02} 
with the explicit symplectic algorithm 
for the Nos\'e-Poincar\'e thermostat by Nos\'e \cite{nose01}. 
We further combine our algorithm with the Andersen barostat \cite{and80} 
to present an explicit symplectic MD algorithm 
for rigid-body molecules in the isothermal-isobaric ensemble. 
An explicit symplectic MD algorithm 
for spherical atoms in the isothermal-isobaric ensemble 
has been presented in references [\onlinecite{okokmd04, okokmd06}]. 
The isothermal-isobaric algorithm in this article is 
an extension of the algorithm for spherical atoms to 
that for rigid-body molecules. 
We also present a symplectic integrator in 
the constant normal pressure and lateral surface area ensemble 
and a symplectic integrator combined with the Parrinello-Rahman algorithm. 

In Section \ref{method:sec} 
we first give brief reviews of the Nos\'e-Poincar\'e thermostat and 
the rigid-body MD algorithm. 
We then explain the symplectic MD algorithms for rigid-body molecules 
in the canonical, isothermal-isobaric, and related ensembles. 
In Section \ref{comp:sec} we compare 
our symplectic MD algorithm 
with nonsymplectic MD algorithms in the canonical ensemble. 
We apply our symplectic MD algorithm to a rigid-body water model 
and make numerical comparisons with the nonsymplectic MD algorithms. 
Section \ref{conc:sec} is devoted to conclusions. 

%
  \section{Methods} \label{method:sec}
%
\subsection{Nos\'e-Poincar\'e thermostat} \label{nose-pcr:sec}
The Nos\'e-Poincar\'e Hamiltonian $H_{\rm NP}$ for $N$ spherical atoms 
at temperature $T_0$ is given by \cite{bond99,nose01} 
\begin{eqnarray}
H_{\rm NP} &=& s 
          \left[
              \sum_{i=1}^N \frac{\bs{p}'^2_i}{2m_is^2} + E(\bs{r}^{\{N\}}) 
            + \frac{P_s^2}{2Q} + g k_{\rm B}T_0 \log s - H_0 
          \right] \nonumber \\
           &=& s \left[ H_{\rm N} (\bs{r}^{\{N\}},\bs{p}'^{\{N\}},s,P_s) 
                      - H_0 \right]~,
           \label{Hnp:eq}
\end{eqnarray}
where 
$\bs{p}'_i$ and $P_s$ are the conjugate momenta for 
the coordinate $\bs{r}_i$ of particle $i$ 
and Nos\'e's additional degree of freedom $s$, respectively. 
We have introduced a simplified notation by the superscript $\{N\}$ 
for the set of coordinate and momentum vectors: 
$\bs{r}^{\{N\}} \equiv (\bs{r}_1, \bs{r}_2, \cdots , \bs{r}_N)^{\rm T}$ and 
$\bs{p}'^{\{N\}} \equiv (\bs{p}'_1, \bs{p}'_2, \cdots , \bs{p}'_N)^{\rm T}$, 
where the superscript $\rm T$ stands for transpose. 
The real momentum $\bs{p}_i$ and the virtual momentum $\bs{p}'_i$ 
are related by 
\begin{equation}
\bs{p}_i = \bs{p}'_i/s. \label{sc1p:eq}
\end{equation}
$E$ is the potential energy. 
The constant $m_i$ is the mass of particle $i$ and 
$Q$ is the artificial ``mass" associated with $s$. 
The constant $g$ corresponds to the number of degrees of freedom. 
In the case of a spherical atomic system, 
$g$ equals $3N$ 
($g$ equals $6N$ in the case of a rigid-body molecular system). 
The Hamiltonian $H_{\rm N}$ is the original Nos\'e Hamiltonian and 
$H_0$ is the initial value of $H_{\rm N}$. 

The equations of motion for the Nos\'e-Poincar\'e thermostat are given by
\begin{eqnarray}
\dot{\bs{r}}_i &=& \frac{\bs{p}_i}{m_i}~, \label{eomnp1:eq} \\
\dot{\bs{p}}_i &=& \bs{F}_i - \frac{\dot{s}}{s} \bs{p}_i~, \label{eomnp2:eq} \\
\dot{s}        &=& s \frac{P_s}{Q}~, \label{eomnp3:eq} \\
\dot{P}_s      &=& \sum_{i=1}^N \frac{\bs{p}^2_i}{m_i} - g k_{\rm B}T_0~,
                   \label{eomnp4:eq}
\end{eqnarray}
where the dot above each variable stands for the time derivative and 
the relation of 
\begin{equation}
H_{\rm N} - H_0 = 0
\end{equation}
is used because $H_{\rm N}$ is conserved. 
Equations (\ref{eomnp1:eq})-(\ref{eomnp4:eq}) 
are the same as those for the Nos\'e thermostat in the real time. 

\subsection{Molecular dynamics algorithm for rigid-body molecules 
in the microcanonical ensemble} 
\label{symp-rg:sec}
Hamiltonian for rigid-body molecules $H_{\rm RB}$ 
are given by \cite{mill02,gold-quat} 
\begin{eqnarray}
H_{\rm RB} &=& \sum_{i=1}^N
                  \frac{1}{8} \bs{\pi}^{\rm T}_i {\stackrel{\leftrightarrow}{\bs S}}(\bs{q}_i) {\stackrel{\leftrightarrow}{\bs D}}_i 
                   {\stackrel{\leftrightarrow}{\bs S}}^{\rm T}(\bs{q}_i) \bs{\pi}_i + E(\bs{q}^{\{N\}})~,
           \label{Hrb:eq}
\end{eqnarray}
where $\bs{q}_i$ is a quaternion of molecule $i$, which indicates 
the orientation of the rigid-body molecule. 
Here, the quaternion ${\bs q} = (q_0, q_1, q_2, q_3)^{\rm T}$ is related to 
the Euler angle ($\phi$, $\theta$, $\psi$) as follows: 
\begin{eqnarray}
 q_0 = \cos \left( \frac{\theta}{2} \right)
       \cos \left( \frac{\phi+\psi}{2} \right)~, \\
 q_1 = \sin \left( \frac{\theta}{2} \right)
       \cos \left( \frac{\phi-\psi}{2} \right)~, \\
 q_2 = \sin \left( \frac{\theta}{2} \right)
       \sin \left( \frac{\phi-\psi}{2} \right)~, \\
 q_3 = \cos \left( \frac{\theta}{2} \right)
       \sin \left( \frac{\phi+\psi}{2} \right)~.
\end{eqnarray}
The elements of the matrix ${\bs S} ({\bs q})$ are given by 
\begin{equation}
{\stackrel{\leftrightarrow}{\bs S}}(\bs{q}) = \left(
         \begin{array}{rrrr}
          q_0 & -q_1 & -q_2 & -q_3 \\ 
          q_1 &  q_0 & -q_3 &  q_2 \\ 
          q_2 &  q_3 &  q_0 & -q_1 \\ 
          q_3 & -q_2 &  q_1 &  q_0 \\ 
         \end{array} 
         \right)~.
\end{equation}
The variable ${\bs \pi}_i$ is the conjugate momentum for ${\bs q}_i$. 
The matrix ${\stackrel{\leftrightarrow}{\bs D}}$ is a 4$\times$4 matrix consisting of the inverse of 
the principal moments of inertia $I_1$, $I_2$, and $I_3$ of molecule $i$: 
\begin{equation}
{\stackrel{\leftrightarrow}{\bs D}} = \left(
         \begin{array}{cccc}
         I^{-1}_0 & 0 & 0 & 0 \\ 
         0 & I^{-1}_1 & 0 & 0 \\ 
         0 & 0 & I^{-1}_2 & 0 \\ 
         0 & 0 & 0 & I^{-1}_3 \\ 
         \end{array} 
         \right)~,
\end{equation}
where $I_0$ is an artificial constant. 
Note that the correct equations of motion 
for rigid-body molecules are obtained 
in the limit of $I_0 \rightarrow \infty$. 
In order to write the equations of motion more elegantly, 
we may introduce the angular velocity 
\begin{eqnarray}
 {\bs \omega}       &=& (  \omega_1,\omega_2,\omega_3)^{\rm T}~, 
\end{eqnarray}
and the four-dimensional angular velocity 
\begin{eqnarray}
 {\bs \omega}^{(4)} &=& (0,\omega_1,\omega_2,\omega_3)^{\rm T}~, 
\end{eqnarray}
where $\omega_1$, $\omega_2$, and $\omega_3$ are the angular velocities 
along each of the corresponding principal axes. 
In the limit of $I_0 \rightarrow \infty$, 
the four-dimensional angular velocity ${\bs \omega}^{(4)}_i$ is 
related to ${\bs \pi}_i$ by 
\begin{equation}
 {\bs \omega}^{(4)}_i = \frac{1}{2} {\stackrel{\leftrightarrow}{\bs D}}_i {\stackrel{\leftrightarrow}{\bs S}}^{\rm T}({\bs q}_i) 
                        {\bs \pi}_i~.
\end{equation}
In this limit the equations of motion for rigid-body molecules 
are obtained as follows: 
\begin{eqnarray}
\dot{\bs{q}}_i &=& 
\frac{1}{2} {\stackrel{\leftrightarrow}{\bs S}}(\bs{q}_i) \bs{\omega}^{(4)}_i~, 
\label{eomr1:eq} \\
{\stackrel{\leftrightarrow}{{\bs I}_i}} \dot{\bs{\omega}}_i &=& 
\bs{N}_i - \bs{\omega}_i \times \left( {\stackrel{\leftrightarrow}{{\bs I}_i}} \bs{\omega}_i \right)~. 
\label{eomr2:eq} 
\end{eqnarray}
Equation (\ref{eomr2:eq}) is called the Euler equation of motion. 
Here, $\bs I$ is the 3$\times$3 diagonal matrix whose diagonal elements are 
$I_1$, $I_2$, and $I_3$. 
The vector ${\bs N}_i$ is the torque of molecule $i$, 
which is calculated by 
\begin{equation}
\bs{N}_i= \sum_{\alpha \in i} \bs{r}_{\alpha} \times \bs{F}_{\alpha}~,
\end{equation}
where $\bs{F}_{\alpha}$ and $\bs{r}_{\alpha}$ are 
the coordinate and force of atom $\alpha$, respectively, 
in a rigid-body-fixed coordinate system for molecule $i$. 
The torque $\bs{N}_i$ is related to the potential energy $E$ by 
\begin{equation}
- \frac{\partial E}{\partial \bs{q}_i} 
= 2 {\stackrel{\leftrightarrow}{\bs S}}(\bs{q}_i) \bs{N}_i^{(4)}~,
\end{equation}
where
\begin{equation}
\bs{N}_i^{(4)}
= \left(\sum_{\alpha \in i} \bs{r}_{\alpha} \cdot  \bs{F}_{\alpha}, 
        \sum_{\alpha \in i} \bs{r}_{\alpha} \times \bs{F}_{\alpha} 
  \right)~. 
\end{equation}

\subsection{Symplectic molecular dynamics algorithm for rigid-body molecules 
combined with the Nos\'e-Poincar\'e thermostat} 
\label{nose-pcr-symp-rg:sec}
We here present the explicit symplectic MD algorithm 
for rigid-body molecules in the canonical ensemble. 
We combine the Nos\'e-Poincar\'e Hamiltonian 
in Eq.~(\ref{Hnp:eq}) \cite{bond99,nose01} and 
the Hamiltonian for rigid-body molecules 
in Eq.~(\ref{Hrb:eq}) \cite{mill02,gold-quat}. 
The Nos\'e-Poincar\'e Hamiltonian for rigid-body molecules is given by 
\begin{eqnarray}
H_{\rm NP-RB} &=& s 
                \left[
                \sum_{i=1}^N \frac{\bs{p}'^2_i}{2m_is^2} + 
                \sum_{i=1}^N \frac{1}{8s^2} 
                       \bs{\pi}'^{\rm T}_i 
                       {\stackrel{\leftrightarrow}{\bs S}}(\bs{q}_i) 
                       {\stackrel{\leftrightarrow}{\bs D}}_i 
                       {\stackrel{\leftrightarrow}{\bs S}}^{\rm T}(\bs{q}_i) 
                       \bs{\pi}'_i 
                \right. \nonumber \\
              & & 
                \left.
                    + E(\bs{r}^{\{N\}}, \bs{q}^{\{N\}}) + 
                \frac{P_s^2}{2Q} + g k_{\rm B}T_0 \log s - H_0 
                \right]~,
                \label{hnprb:eq}
\end{eqnarray}
where $\bs{r}^{\{N\}}=({\bs r}_1, {\bs r}_2, \cdots, {\bs r}_N)^{\rm T}$ 
stands for the set of the coordinates of 
the center of mass for the rigid-body molecules. 
The vector $\bs{\pi}'_i$ is the conjugate momentum 
for quaternion ${\bs q}_i$. 
The real momentum ${\bs \pi}_i$ of the quaternion is related to 
the virtual momentum ${\bs \pi}'_i$ by 
\begin{eqnarray}
 {\bs \pi}_i &=& \frac{{\bs \pi}'_i}{s}~. 
\end{eqnarray}
The equations of motion are given from 
the Hamiltonian in Eq.~(\ref{hnprb:eq}) by 
\begin{eqnarray}
\dot{\bs{r}}_i &=& \frac{\bs{p}_i}{m_i}~, \label{eomnpr1:eq} \\
\dot{\bs{p}}_i &=& \bs{F}_i - \frac{\dot{s}}{s} \bs{p}_i~, 
                   \label{eomnpr2:eq} \\
\dot{\bs{q}}_i &=& \frac{1}{2} {\stackrel{\leftrightarrow}{\bs S}}(\bs{q}_i) 
                   \bs{\omega}^{(4)}_i~,
                   \label{eomnpr3:eq} \\
{\stackrel{\leftrightarrow}{{\bs I}_i}} \dot{\bs{\omega}}_i &=& 
                   \bs{N}_i - \bs{\omega}_i 
                   \times \left( {\stackrel{\leftrightarrow}{{\bs I}_i}} 
                   \bs{\omega}_i \right)
                 - \frac{\dot{s}}{s}{\stackrel{\leftrightarrow}{{\bs I}_i}} 
                   \bs{\omega}_i~, 
                   \label{eomnpr4:eq} \\
\dot{s}        &=& s \frac{P_s}{Q}~, \label{eomnpr5:eq} \\
\dot{P}_s      &=& \sum_{i=1}^N \frac{\bs{p}^2_i}{m_i} 
                 + \sum_{i=1}^N \bs{\omega}^{\rm T}_i 
                   {\stackrel{\leftrightarrow}{{\bs I}_i}} \bs{\omega}_i 
                 - g k_{\rm B}T_0~.
                   \label{eomnpr6:eq}
\end{eqnarray}

The time development of a physical quantity $Z(\bs{\Gamma})$ in the phase 
space $\bs{\Gamma} \equiv (\bs{r}^{\{N\}},\bs{p}'^{\{N\}},\bs{q}^{\{N\}},\bs{\pi}'^{\{N\}},s,P_s)^{\rm T}$ 
is written by 
\begin{equation}
\frac{dZ}{dt} = \dot{\bs{\Gamma}} \cdot 
\frac{\partial Z}{\partial \bs{\Gamma}}~.
\end{equation}
The formal solution of the time development of $Z$ 
from time $t$ to $t+\Delta t$ is given by 
\begin{equation}
Z(t+\Delta t) = {\rm e}^{D \Delta t} Z(t)~,
\end{equation}
where ${\rm e}^{D \Delta t}$ is called a time propagator. 
The operator $D$ is defined by 
\begin{equation}
D \equiv \dot{\bs{\Gamma}} \cdot \frac{\partial}{\partial \bs{\Gamma}}~.
\label{D:eq}
\end{equation}

In the symplectic algorithm, 
the Hamiltonian in Eq.~(\ref{hnprb:eq}) is separated 
into six terms here as follows: 
\begin{eqnarray}
H_{\rm NP-RB } &=& H_{\rm NP-RB0} + H_{\rm NP-RB1} + H_{\rm NP-RB2} 
                   \nonumber \\
               &+& H_{\rm NP-RB3} + H_{\rm NP-RB4} + H_{\rm NP-RB5}~, \\
H_{\rm NP-RB0} &=& s \sum_{i=1}^N \frac{1}{8I_0 s^2} 
                 \left( \bs{\pi}'^{\rm T}_i 
                 {\stackrel{\leftrightarrow}{\bcal P}}_0 \bs{q}_i 
                 \right)^2~, \\
H_{\rm NP-RB1} &=& s \left[\sum_{i=1}^N \frac{\bs{p}'^2_i}{2m_is^2}
                   + \sum_{i=1}^N \frac{1}{8I_1 s^2} 
                       \left( \bs{\pi}'^{\rm T}_i 
                       {\stackrel{\leftrightarrow}{\bcal P}}_1 \bs{q}_i 
                       \right)^2
                   + g k_{\rm B}T_0 \log s - H_0 
                 \right]~, \\
H_{\rm NP-RB2} &=& s \sum_{i=1}^N \frac{1}{8I_2 s^2} 
                 \left( \bs{\pi}'^{\rm T}_i 
                 {\stackrel{\leftrightarrow}{\bcal P}}_2 \bs{q}_i 
                 \right)^2~, \\
H_{\rm NP-RB3} &=& s \sum_{i=1}^N \frac{1}{8I_3 s^2} 
                 \left( \bs{\pi}'^{\rm T}_i 
                 {\stackrel{\leftrightarrow}{\bcal P}}_3 \bs{q}_i 
                 \right)^2~, \\
H_{\rm NP-RB4} &=& s E(\bs{r}^{\{N\}}, \bs{q}^{\{N\}})~, \\
H_{\rm NP-RB5} &=& s \frac{P_s^2}{2Q}~,
\end{eqnarray}
where 
\begin{eqnarray}
{\stackrel{\leftrightarrow}{\bcal P}}_0 \bs{q} &=& 
(\ \ q_0, \ \ q_1, \ \ q_2, \ \ q_3)^{\rm T}, \\
{\stackrel{\leftrightarrow}{\bcal P}}_1 \bs{q} &=& 
(-q_1, \ \ q_0, \ \ q_3,-q_2)^{\rm T}, \\
{\stackrel{\leftrightarrow}{\bcal P}}_2 \bs{q} &=& 
(-q_2,-q_3, \ \ q_0, \ \ q_1)^{\rm T}, \\
{\stackrel{\leftrightarrow}{\bcal P}}_3 \bs{q} &=& 
(-q_3, \ \ q_2,-q_1, \ \ q_0)^{\rm T}. 
\end{eqnarray}
In the limit of $I_0 \rightarrow \infty$, $H_{\rm NP-RB0}$ goes to zero: 
$H_{\rm NP-RB0} \rightarrow 0$. 
Hereafter, then only Hamiltonians 
from $H_{\rm NP-RB1}$ to $H_{\rm NP-RB5}$ are considered. 
The second-order formula with respect to $\Delta t$ 
is obtained by the decomposition 
of the time propagator ${\rm exp} \left[ D \Delta t \right]$ into 
a product of five time propagators: 
\begin{eqnarray}
{\rm exp} \left[ D \Delta t \right] 
&=&
{\rm exp} \left[ D_5 \frac{\Delta t}{2} \right]
{\rm exp} \left[ D_4 \frac{\Delta t}{2} \right]
{\rm exp} \left[ D_3 \frac{\Delta t}{2} \right]
{\rm exp} \left[ D_2 \frac{\Delta t}{2} \right]
{\rm exp} \left[ D_1 \Delta t \right] \nonumber \\
&\times&
{\rm exp} \left[ D_2 \frac{\Delta t}{2} \right]
{\rm exp} \left[ D_3 \frac{\Delta t}{2} \right]
{\rm exp} \left[ D_4 \frac{\Delta t}{2} \right]
{\rm exp} \left[ D_5 \frac{\Delta t}{2} \right] \nonumber \\
&+& O \left( \left( \Delta t \right)^3 \right)~.
\label{dcmp1:eq}
\end{eqnarray}
Higher-order formulae can also be obtained in a similar manner. 
The explicit form of each operator is as follows: 
\begin{eqnarray}
D_1
&=&
\sum_{i=1}^N \left(
  \frac{\partial H_{\rm NP-RB1}}{\partial \bs{p}'_i} \cdot
  \frac{\partial }{\partial \bs{r}_i}
- \frac{\partial H_{\rm NP-RB1}}{\partial \bs{r}_i} \cdot
  \frac{\partial }{\partial \bs{p}'_i}
\right) \nonumber \\
&+&
\sum_{i=1}^N \left(
  \frac{\partial H_{\rm NP-RB1}}{\partial \bs{\pi}_i} \cdot
  \frac{\partial }{\partial \bs{q}_i}
- \frac{\partial H_{\rm NP-RB1}}{\partial \bs{q}_i} \cdot
  \frac{\partial }{\partial \bs{\pi}_i}
\right) \nonumber \\
&+&
  \frac{\partial H_{\rm NP-RB1}}{\partial P_s}
  \frac{\partial }{\partial s}
- \frac{\partial H_{\rm NP-RB1}}{\partial s}
  \frac{\partial }{\partial P_s} \nonumber \\
&=&
\sum_{i=1}^N \frac{\bs{p}'_i}{m_is} \cdot \frac{\partial }{\partial \bs{r}_i} 
\nonumber \\
&+&
\sum_{i=1}^N \frac{1}{4I_1 s} 
       \left( \bs{\pi}'^{\rm T}_i {\stackrel{\leftrightarrow}{\bcal P}}_1 \bs{q}_i \right) 
       \left( {\stackrel{\leftrightarrow}{\bcal P}}_1 \bs{q}_i \right) \cdot 
\frac{\partial }{\partial \bs{q}_i} 
+
\sum_{i=1}^N \frac{1}{4I_1 s} 
       \left( \bs{\pi}'^{\rm T}_i {\stackrel{\leftrightarrow}{\bcal P}}_1 \bs{q}_i \right) 
       \left( {\stackrel{\leftrightarrow}{\bcal P}}_1 \bs{\pi}'_i \right) \cdot 
\frac{\partial }{\partial \bs{\pi}'_i} \nonumber \\
&+&
\left[ \sum_{i=1}^N \frac{\bs{p}'^2_i}{2m_is^2} + 
       \sum_{i=1}^N \frac{1}{8I_1 s^2} 
              \left( \bs{\pi}'^{\rm T}_i {\stackrel{\leftrightarrow}{\bcal P}}_1 \bs{q}_i \right)^2
     - g k_{\rm B}T_0 \log s + H_0 - g k_{\rm B}T_0
\right]
\frac{\partial }{\partial P_s}~, \nonumber \\
\\
D_2
&=&
\sum_{i=1}^N \frac{1}{4I_2 s} 
       \left( \bs{\pi}'^{\rm T}_i {\stackrel{\leftrightarrow}{\bcal P}}_2 \bs{q}_i \right) 
       \left( {\stackrel{\leftrightarrow}{\bcal P}}_2 \bs{q}_i \right) \cdot 
\frac{\partial }{\partial \bs{q}_i} 
+
\sum_{i=1}^N \frac{1}{4I_2 s} 
       \left( \bs{\pi}'^{\rm T}_i {\stackrel{\leftrightarrow}{\bcal P}}_2 \bs{q}_i \right) 
       \left( {\stackrel{\leftrightarrow}{\bcal P}}_2 \bs{\pi}'_i \right) \cdot 
\frac{\partial }{\partial \bs{\pi}'_i} \nonumber \\
&+&
\left[ \sum_{i=1}^N \frac{1}{8I_2 s^2} 
       \left( \bs{\pi}'^{\rm T}_i {\stackrel{\leftrightarrow}{\bcal P}}_2 \bs{q}_i \right)^2
\right]
\frac{\partial }{\partial P_s}~,
\\
D_3
&=&
\sum_{i=1}^N \frac{1}{4I_3 s} 
       \left( \bs{\pi}'^{\rm T}_i {\stackrel{\leftrightarrow}{\bcal P}}_3 \bs{q}_i \right) 
       \left( {\stackrel{\leftrightarrow}{\bcal P}}_3 \bs{q}_i \right) \cdot 
\frac{\partial }{\partial \bs{q}_i} 
+
\sum_{i=1}^N \frac{1}{4I_3 s} 
       \left( \bs{\pi}'^{\rm T}_i {\stackrel{\leftrightarrow}{\bcal P}}_3 \bs{q}_i \right) 
       \left( {\stackrel{\leftrightarrow}{\bcal P}}_3 \bs{\pi}'_i \right) \cdot 
\frac{\partial }{\partial \bs{\pi}'_i} \nonumber \\
&+&
\left[ \sum_{i=1}^N \frac{1}{8I_3 s^2} 
       \left( \bs{\pi}'^{\rm T}_i {\stackrel{\leftrightarrow}{\bcal P}}_3 \bs{q}_i \right)^2
\right]
\frac{\partial }{\partial P_s}~,
\\
D_4
&=&
\sum_{i=1}^N s \bs{F}_i \cdot \frac{\partial }{\partial \bs{p}'_i} 
+
\sum_{i=1}^N 2s \left( {\stackrel{\leftrightarrow}{\bs S}}(\bs{q}_i) \bs{N}^{(4)}_i \right) 
\cdot
\frac{\partial }{\partial \bs{\pi}'_i} 
- E(\bs{r}^{\{N\}}, \bs{q}^{\{N\}})
\frac{\partial }{\partial P_s}~,
\\
D_5
&=&
  \frac{s P_s}{Q} \frac{\partial }{\partial s} 
- \frac{P^2_s}{2Q} \frac{\partial }{\partial P_s}~.
\end{eqnarray}
There is no term higher than the second power of $\Delta t$ 
in the time developments by $D_4$, 
because there is no conjugate pair in $H_{\rm NP-RB4}$. 
Although there is a conjugate pair of 
$\bs{q}_i$ and $\bs{\pi}'_i$ in $H_{\rm NP-RB1}$, 
the time developments of $\bs{q}_i$ and $\bs{\pi}'_i$ 
by $H_{\rm NP-RB1}$ are given by \cite{mill02}
\begin{eqnarray}
{\rm exp} \left[ D_1 \Delta t \right] \bs{q}_i 
&=&
\cos \left( \zeta_{i1} \Delta t \right) \bs{q}_i + 
\sin \left( \zeta_{i1} \Delta t \right) {\stackrel{\leftrightarrow}{\bcal P}}_1 \bs{q}_i~, \\
{\rm exp} \left[ D_1 \Delta t \right] \bs{\pi}'_i 
&=&
\cos \left( \zeta_{i1} \Delta t \right) \bs{\pi}'_i + 
\sin \left( \zeta_{i1} \Delta t \right) {\stackrel{\leftrightarrow}{\bcal P}}_1 \bs{\pi}'_i~,
\end{eqnarray}
where
\begin{equation}
\zeta_{i1} = \frac{1}{4I_1 s} \bs{\pi}'^{\rm T}_i {\stackrel{\leftrightarrow}{\bcal P}}_1 \bs{q}_i~.
\end{equation}
The time developments of $\bs{q}_i$ and $\bs{\pi}'_i$ 
by $D_2$ and $D_3$ are also obtained in the same way. 
Although there is another conjugate pair of $s$ and $P_s$ in $H_{\rm NP-RB5}$, 
the time developments of $s$ and $P_s$ by $D_5$ are given explicitly 
by \cite{nose01}
\begin{eqnarray}
{\rm exp} \left[ D_5 \Delta t \right] s 
&=&
s \left( 1+ \frac{P_s}{2Q} \Delta t \right)^2~, \\
{\rm exp} \left[ D_5 \Delta t \right] P_s 
&=&
P_s \bigg/ \left( 1+ \frac{P_s}{2Q} \Delta t \right)~.
\end{eqnarray}

Finally, the explicit symplectic time developments 
for rigid-body molecules in the canonical ensemble 
is obtained from Eq.~(\ref{dcmp1:eq}). 
Here, a symbol of $\leftarrow$ stands for a substitution 
in a computational program 
(i.e., the variables in each step adopt 
the substitutions in the preceding steps): \\
Step 1. ${\rm exp} \left[ D_5 \Delta t /2 \right]$ operation: 
\begin{eqnarray}
s 
&\leftarrow&
s \left( 1+ \frac{P_s}{2Q}\frac{\Delta t}{2} \right)^2~,
\label{intss-1:eq}
\\
P_s 
&\leftarrow&
P_s \bigg/ \left( 1+ \frac{P_s}{2Q}\frac{\Delta t}{2} \right)~.
\end{eqnarray}
Step 2. ${\rm exp} \left[ D_4 \Delta t /2 \right]$ operation: 
\begin{eqnarray}
\bs{p}'_i
&\leftarrow&
\bs{p}'_i + s \bs{F}_i \frac{\Delta t}{2}~,
\\
\bs{\pi}'_i
&\leftarrow&
\bs{\pi}'_i + 2s {\stackrel{\leftrightarrow}{\bs S}}(\bs{q}_i) \bs{N}^{(4)}_i \frac{\Delta t}{2}~,
\\
P_s 
&\leftarrow&
P_s - E(\bs{r}^{\{N\}}, \bs{q}^{\{N\}}) \frac{\Delta t}{2}~.
\end{eqnarray}
Step 3. ${\rm exp} \left[ D_3 \Delta t /2 \right]$ operation: 
\begin{eqnarray}
\zeta_{i3} 
&\leftarrow&
\frac{1}{4I_3 s} \bs{\pi}'^{\rm T}_i 
{\stackrel{\leftrightarrow}{\bcal P}}_3 \bs{q}_i~,
\\
\bs{q}_i 
&\leftarrow&
\cos \left( \zeta_{i3} \frac{\Delta t}{2} \right) \bs{q}_i + 
\sin \left( \zeta_{i3} \frac{\Delta t}{2} \right) 
{\stackrel{\leftrightarrow}{\bcal P}}_3 \bs{q}_i~,
\\
\bs{\pi}'_i 
&\leftarrow&
\cos \left( \zeta_{i3} \frac{\Delta t}{2} \right) \bs{\pi}'_i + 
\sin \left( \zeta_{i3} \frac{\Delta t}{2} \right) 
{\stackrel{\leftrightarrow}{\bcal P}}_3 \bs{\pi}'_i~,
\\
P_s 
&\leftarrow&
P_s + 
\left( \sum_{i=1}^N 2I_3 \zeta^2_{i3} \right) 
\frac{\Delta t}{2}~.
\end{eqnarray}
Step 4. ${\rm exp} \left[ D_2 \Delta t /2 \right]$ operation: 
\begin{eqnarray}
\zeta_{i2} 
&\leftarrow&
\frac{1}{4I_2 s} \bs{\pi}'^{\rm T}_i 
{\stackrel{\leftrightarrow}{\bcal P}}_2 \bs{q}_i~,
\\
\bs{q}_i 
&\leftarrow&
\cos \left( \zeta_{i2} \frac{\Delta t}{2} \right) \bs{q}_i + 
\sin \left( \zeta_{i2} \frac{\Delta t}{2} \right) 
{\stackrel{\leftrightarrow}{\bcal P}}_2 \bs{q}_i~,
\\
\bs{\pi}'_i 
&\leftarrow&
\cos \left( \zeta_{i2} \frac{\Delta t}{2} \right) \bs{\pi}'_i + 
\sin \left( \zeta_{i2} \frac{\Delta t}{2} \right) 
{\stackrel{\leftrightarrow}{\bcal P}}_2 \bs{\pi}'_i~,
\\
P_s 
&\leftarrow&
P_s + 
\left( \sum_{i=1}^N 2I_2 \zeta^2_{i2} \right) 
\frac{\Delta t}{2}~.
\end{eqnarray}
Step 5. ${\rm exp} \left[ D_1 \Delta t \right]$ operation: 
\begin{eqnarray}
\bs{r}_i 
&\leftarrow&
\bs{r}_i + \frac{\bs{p}'_i}{m_is} \Delta t~,
\\
\zeta_{i1} 
&\leftarrow&
\frac{1}{4I_1 s} \bs{\pi}'^{\rm T}_i 
{\stackrel{\leftrightarrow}{\bcal P}}_1 \bs{q}_i~,
\\
\bs{q}_i 
&\leftarrow&
\cos \left( \zeta_{i1} \Delta t \right) \bs{q}_i + 
\sin \left( \zeta_{i1} \Delta t \right) 
{\stackrel{\leftrightarrow}{\bcal P}}_1 \bs{q}_i~,
\\
\bs{\pi}'_i 
&\leftarrow&
\cos \left( \zeta_{i1} \Delta t \right) \bs{\pi}'_i + 
\sin \left( \zeta_{i1} \Delta t \right) 
{\stackrel{\leftrightarrow}{\bcal P}}_1 \bs{\pi}'_i~,
\\
P_s 
&\leftarrow&
P_s + 
\left( \sum_{i=1}^N \frac{\bs{p}'^2_i}{2m_is^2} + 
       \sum_{i=1}^N 2I_1 \zeta^2_{i1} 
     - g k_{\rm B}T_0 \log s + H_0 - g k_{\rm B}T_0
\right) \Delta t~.
\end{eqnarray}
Step 6. ${\rm exp} \left[ D_2 \Delta t /2 \right]$ operation: 
\begin{eqnarray}
\zeta_{i2} 
&\leftarrow&
\frac{1}{4I_2 s} \bs{\pi}'^{\rm T}_i 
{\stackrel{\leftrightarrow}{\bcal P}}_2 \bs{q}_i~,
\\
\bs{q}_i 
&\leftarrow&
\cos \left( \zeta_{i2} \frac{\Delta t}{2} \right) \bs{q}_i + 
\sin \left( \zeta_{i2} \frac{\Delta t}{2} \right) 
{\stackrel{\leftrightarrow}{\bcal P}}_2 \bs{q}_i~,
\\
\bs{\pi}'_i 
&\leftarrow&
\cos \left( \zeta_{i2} \frac{\Delta t}{2} \right) \bs{\pi}'_i + 
\sin \left( \zeta_{i2} \frac{\Delta t}{2} \right) 
{\stackrel{\leftrightarrow}{\bcal P}}_2 \bs{\pi}'_i~,
\\
P_s 
&\leftarrow&
P_s + 
\left( \sum_{i=1}^N 2I_2 \zeta^2_{i2} \right) 
\frac{\Delta t}{2}~.
\end{eqnarray}
Step 7. ${\rm exp} \left[ D_3 \Delta t /2 \right]$ operation: 
\begin{eqnarray}
\zeta_{i3} 
&\leftarrow&
\frac{1}{4I_3 s} \bs{\pi}'^{\rm T}_i 
{\stackrel{\leftrightarrow}{\bcal P}}_3 \bs{q}_i~,
\\
\bs{q}_i 
&\leftarrow&
\cos \left( \zeta_{i3} \frac{\Delta t}{2} \right) \bs{q}_i + 
\sin \left( \zeta_{i3} \frac{\Delta t}{2} \right) 
{\stackrel{\leftrightarrow}{\bcal P}}_3 \bs{q}_i~,
\\
\bs{\pi}'_i 
&\leftarrow&
\cos \left( \zeta_{i3} \frac{\Delta t}{2} \right) \bs{\pi}'_i + 
\sin \left( \zeta_{i3} \frac{\Delta t}{2} \right) 
{\stackrel{\leftrightarrow}{\bcal P}}_3 \bs{\pi}'_i~,
\\
P_s 
&\leftarrow&
P_s + 
\left( \sum_{i=1}^N 2I_3 \zeta^2_{i3} \right) 
\frac{\Delta t}{2}~.
\end{eqnarray}
Step 8. ${\rm exp} \left[ D_4 \Delta t /2 \right]$ operation: 
\begin{eqnarray}
\bs{p}'_i
&\leftarrow&
\bs{p}'_i + s \bs{F}_i \frac{\Delta t}{2}~,
\\
\bs{\pi}'_i
&\leftarrow&
\bs{\pi}'_i + 2s {\stackrel{\leftrightarrow}{\bs S}}(\bs{q}_i) 
\bs{N}^{(4)}_i \frac{\Delta t}{2}~,
\\
P_s 
&\leftarrow&
P_s - E(\bs{r}^{\{N\}}, \bs{q}^{\{N\}}) \frac{\Delta t}{2}~.
\end{eqnarray}
Step 9. ${\rm exp} \left[ D_5 \Delta t /2 \right]$ operation: 
\begin{eqnarray}
s 
&\leftarrow&
s \left( 1+ \frac{P_s}{2Q}\frac{\Delta t}{2} \right)^2~,
\\
P_s 
&\leftarrow&
P_s \bigg/ \left( 1+ \frac{P_s}{2Q}\frac{\Delta t}{2} \right)~.
\label{intss-f:eq}
\end{eqnarray}

\subsection{Symplectic molecular dynamics algorithm for rigid-body molecules 
combined with the Nos\'e-Poincar\'e thermostat and the Andersen barostat} 
\label{nose-pcr-and-symp-rg:sec}
In this subsection we present the explicit symplectic MD algorithm 
for rigid-body molecules in the isothermal-isobaric ensemble. 
Hamiltonian for rigid-body molecules 
at temperature $T_0$ and pressure $P_0$ 
is given by combing the Hamiltonian in Eq.~(\ref{hnprb:eq}) and 
the Andersen barostat \cite{and80} as follows: 
\begin{eqnarray}
H_{\rm NPA-RB} &=& s 
                \left[
                \sum_{i=1}^N \frac{\tilde{\bs{p}}^2_i}
                                  {2m_is^2V^{\frac{2}{3}}} + 
                \sum_{i=1}^N \frac{1}{8s^2} 
                       \bs{\pi}'^{\rm T}_i 
                       {\stackrel{\leftrightarrow}{\bs S}}(\bs{q}_i) 
                       {\stackrel{\leftrightarrow}{\bs D}}_i 
                       {\stackrel{\leftrightarrow}{\bs S}}^{\rm T}(\bs{q}_i) 
                       \bs{\pi}'_i 
                       + E(\tilde{\bs{r}}^{\{N\}}, \bs{q}^{\{N\}}, V) 
                \right.
                \nonumber \\
            & & \left. + 
                \frac{P_s^2}{2Q} + g k_{\rm B}T_0 \log s + 
                \frac{P_V^2}{2W} + P_0V - H_0 
                \right]
                \nonumber \\
            &=& s \left[
                       H_{\rm NA}
                       (\tilde{\bs{r}}^{\{N\}},\tilde{\bs{p}}^{\{N\}},
                        \bs{q}^{\{N\}},\bs{\pi}'^{\{N\}},
                        s,P_s,V,P_V) - H_0 
                  \right]~,
\end{eqnarray}
where $\tilde{\bs{p}}_i$ and $\tilde{\bs{r}}_i$ are the scaled momentum and 
the scaled coordinate by volume $V$ and 
the degree of the Nos\'e-Poincar\'e thermostat $s$. 
They are related to $\bs{p}_i$ and $\bs{r}_i$ by 
\begin{eqnarray}
\bs{p}_i &=& \tilde{\bs{p}}_i/sV^{\frac{1}{3}}~, \label{sc2p:eq} \\
\bs{r}_i &=& V^{\frac{1}{3}} \tilde{\bs{r}}_i~. \label{sc2r:eq} 
\end{eqnarray}
The constant $W$ is the ``mass" associated with $V$. 
The variable $P_V$ is the conjugate momenta for $V$. 
The constant $H_0$ here is the initial value of the 
Nos\'e-Andersen Hamiltonian $H_{\rm NA}$. 
The equations of motion are given by 
\begin{eqnarray}
\dot{\bs{r}}_i &=& \frac{\bs{p}_i}{m_i} 
                 + \frac{\dot{V}}{3V}\bs{r}_i~, \label{eomnpar1:eq} \\
\dot{\bs{p}}_i &=& \bs{F}_i 
                 - \left( \frac{\dot{s}}{s} + \frac{\dot{V}}{3V} \right) 
                   \bs{p}_i~, \label{eomnpar2:eq} \\
\dot{\bs{q}}_i &=& \frac{1}{2} {\stackrel{\leftrightarrow}{\bs S}}(\bs{q}_i) 
                   \bs{\omega}^{(4)}_i~, 
                   \label{eomnpar3:eq} \\
{\stackrel{\leftrightarrow}{{\bs I}_i}} \dot{\bs{\omega}}_i &=& 
                   \bs{N}_i - \bs{\omega}_i \times 
                   \left( {\stackrel{\leftrightarrow}{{\bs I}_i}} 
                   \bs{\omega}_i \right)
                 - \frac{\dot{s}}{s}
                   {\stackrel{\leftrightarrow}{{\bs I}_i}} 
                   \bs{\omega}_i~, 
                   \label{eomnpar4:eq} \\
\dot{s}        &=& s \frac{P_s}{Q}~, \label{eomnpar5:eq} \\
\dot{P}_s      &=& \sum_{i=1}^N \frac{\bs{p}^2_i}{m_i} 
                 + \sum_{i=1}^N \bs{\omega}^{\rm T}_i 
                   {\stackrel{\leftrightarrow}{{\bs I}_i}} \bs{\omega}_i 
                 - g k_{\rm B}T_0~, 
                   \label{eomnpar6:eq} \\
\dot{V}        &=& s \frac{P_V}{W}~, \label{eomnpar7:eq} \\
\dot{P}_V      &=& s \left[ \frac{1}{3V} 
                     \left( \sum_{i=1}^N \frac{\bs{p}^2_i}{m_i} 
                          + \sum_{i=1}^N \bs{F}_i \cdot \bs{r}_i 
                     \right)
                 - P_0 \right]~.
                   \label{eomnpar8:eq}
\end{eqnarray}
where the relation of 
\begin{equation}
H_{\rm NA} - H_0 = 0
\end{equation}
is used. 

The Hamiltonian in the isothermal-isobaric ensemble 
is separated into six terms as follows: 
\begin{eqnarray}
H_{\rm NPA-RB } &=& H_{\rm NPA-RB1} + H_{\rm NPA-RB2} + H_{\rm NPA-RB3} 
                    \nonumber \\
                &+& H_{\rm NPA-RB4} + H_{\rm NPA-RB5} + H_{\rm NPA-RB6}~, \\
H_{\rm NPA-RB1} &=& s \left[\sum_{i=1}^N \frac{\tilde{\bs{p}}^2_i}
                                        {2m_is^2V^{\frac{2}{3}}}
                   + \sum_{i=1}^N \frac{1}{8I_1 s^2} 
                     \left( \bs{\pi}'^{\rm T}_i 
                     {\stackrel{\leftrightarrow}{\bcal P}}_1 
                     \bs{q}_i \right)^2
                   + g k_{\rm B}T_0 \log s - H_0 
                 \right]~, \\
H_{\rm NPA-RB2} &=& s \sum_{i=1}^N \frac{1}{8I_2 s^2} 
                 \left( \bs{\pi}'^{\rm T}_i 
                 {\stackrel{\leftrightarrow}{\bcal P}}_2 \bs{q}_i \right)^2~, \\
H_{\rm NPA-RB3} &=& s \sum_{i=1}^N \frac{1}{8I_3 s^2} 
                 \left( \bs{\pi}'^{\rm T}_i 
                 {\stackrel{\leftrightarrow}{\bcal P}}_3 \bs{q}_i \right)^2~, \\
H_{\rm NPA-RB4} &=& s \frac{P_V^2}{2W}~, \\
H_{\rm NPA-RB5} &=& s \left[ E(\tilde{\bs{r}}^{\{N\}}, \bs{q}^{\{N\}}, V) 
                  + P_0V \right]~, \\
H_{\rm NPA-RB6} &=& s \frac{P_s^2}{2Q}~,
\end{eqnarray}
where the term of $s \sum_{i=1}^N \left( \bs{\pi}'^{\rm T}_i {\stackrel{\leftrightarrow}{\bcal P}}_0 \bs{q}_i \right)^2 / {8I_0 s^2}$ 
has been neglected again because it is zero 
in the limit of $I_0 \rightarrow \infty$. 
As in the decomposition in Eq.~(\ref{dcmp1:eq}) in the canonical ensemble, 
the second-order formula is obtained 
for the time propagator ${\rm exp} \left[ D \Delta t \right]$ as 
a product of six time propagators: 
\begin{eqnarray}
{\rm exp} \left[ D \Delta t \right] 
&=&
{\rm exp} \left[ D_6 \frac{\Delta t}{2} \right]
{\rm exp} \left[ D_5 \frac{\Delta t}{2} \right]
{\rm exp} \left[ D_4 \frac{\Delta t}{2} \right]
{\rm exp} \left[ D_3 \frac{\Delta t}{2} \right] \nonumber \\
&\times&
{\rm exp} \left[ D_2 \frac{\Delta t}{2} \right]
{\rm exp} \left[ D_1 \Delta t \right] 
{\rm exp} \left[ D_2 \frac{\Delta t}{2} \right] \nonumber \\
&\times&
{\rm exp} \left[ D_3 \frac{\Delta t}{2} \right]
{\rm exp} \left[ D_4 \frac{\Delta t}{2} \right]
{\rm exp} \left[ D_5 \frac{\Delta t}{2} \right]
{\rm exp} \left[ D_6 \frac{\Delta t}{2} \right] \nonumber \\
&+& O \left( \left( \Delta t \right)^3 \right)~,
\label{dcmp2:eq}
\end{eqnarray}
where $D_1$, $D_2$, $\cdots$ $D_6$ are the time propagators 
which correspond to $H_{\rm NPA-RB1}$, $H_{\rm NPA-RB2}$, 
$\cdots$, $H_{\rm NPA-RB6}$, respectively.

According to the decomposition in Eq.~(\ref{dcmp2:eq}), 
the explicit symplectic time developments 
for rigid-body molecules in the isothermal-isobaric ensemble 
are given as follows: \\
Step 1. ${\rm exp} \left[ D_6 \Delta t /2 \right]$ operation: 
\begin{eqnarray}
s 
&\leftarrow&
s \left( 1+ \frac{P_s}{2Q}\frac{\Delta t}{2} \right)^2~,
\\
P_s 
&\leftarrow&
P_s \bigg/ \left( 1+ \frac{P_s}{2Q}\frac{\Delta t}{2} \right)~.
\end{eqnarray}
Step 2. ${\rm exp} \left[ D_5 \Delta t /2 \right]$ operation: 
\begin{eqnarray}
\tilde{\bs{p}}_i
&\leftarrow&
\tilde{\bs{p}}_i + s V^{\frac{1}{3}} \bs{F}_i \frac{\Delta t}{2}~,
\\
\bs{\pi}'_i
&\leftarrow&
\bs{\pi}'_i + 2s {\stackrel{\leftrightarrow}{\bs S}}(\bs{q}_i) \bs{N}^{(4)}_i \frac{\Delta t}{2}~,
\\
P_s 
&\leftarrow&
P_s - \left[ E(\tilde{\bs{r}}^{\{N\}}, \bs{q}^{\{N\}}, V) 
+ P_0V \right] \frac{\Delta t}{2}~,
\\
P_V 
&\leftarrow&
P_V 
+ s \left( \frac{1}{3V} \sum_{i=1}^N \bs{F}_i \cdot \bs{r}_i - P_0 \right) 
    \frac{\Delta t}{2}~.
\end{eqnarray}
Step 3. ${\rm exp} \left[ D_4 \Delta t /2 \right]$ operation: 
\begin{eqnarray}
P_s 
&\leftarrow&
P_s - \frac{P^2_V}{2W} \frac{\Delta t}{2}~,
\\
V 
&\leftarrow&
V + s \frac{P_V}{W} \frac{\Delta t}{2}~.
\end{eqnarray}
Step 4. ${\rm exp} \left[ D_3 \Delta t /2 \right]$ operation: 
\begin{eqnarray}
\zeta_{i3} 
&\leftarrow&
\frac{1}{4I_3 s} \bs{\pi}'^{\rm T}_i {\stackrel{\leftrightarrow}{\bcal P}}_3 \bs{q}_i~,
\\
\bs{q}_i 
&\leftarrow&
\cos \left( \zeta_{i3} \frac{\Delta t}{2} \right) \bs{q}_i + 
\sin \left( \zeta_{i3} \frac{\Delta t}{2} \right) {\stackrel{\leftrightarrow}{\bcal P}}_3 \bs{q}_i~,
\\
\bs{\pi}'_i 
&\leftarrow&
\cos \left( \zeta_{i3} \frac{\Delta t}{2} \right) \bs{\pi}'_i + 
\sin \left( \zeta_{i3} \frac{\Delta t}{2} \right) {\stackrel{\leftrightarrow}{\bcal P}}_3 \bs{\pi}'_i~,
\\
P_s 
&\leftarrow&
P_s + 
\left( \sum_{i=1}^N 2I_3 \zeta^2_{i3} \right) 
\frac{\Delta t}{2}~.
\end{eqnarray}
Step 5. ${\rm exp} \left[ D_2 \Delta t /2 \right]$ operation: 
\begin{eqnarray}
\zeta_{i2} 
&\leftarrow&
\frac{1}{4I_2 s} \bs{\pi}'^{\rm T}_i {\stackrel{\leftrightarrow}{\bcal P}}_2 \bs{q}_i~,
\\
\bs{q}_i 
&\leftarrow&
\cos \left( \zeta_{i2} \frac{\Delta t}{2} \right) \bs{q}_i + 
\sin \left( \zeta_{i2} \frac{\Delta t}{2} \right) {\stackrel{\leftrightarrow}{\bcal P}}_2 \bs{q}_i~,
\\
\bs{\pi}'_i 
&\leftarrow&
\cos \left( \zeta_{i2} \frac{\Delta t}{2} \right) \bs{\pi}'_i + 
\sin \left( \zeta_{i2} \frac{\Delta t}{2} \right) {\stackrel{\leftrightarrow}{\bcal P}}_2 \bs{\pi}'_i~,
\\
P_s 
&\leftarrow&
P_s + 
\left( \sum_{i=1}^N 2I_2 \zeta^2_{i2} \right) 
\frac{\Delta t}{2}~.
\end{eqnarray}
Step 6. ${\rm exp} \left[ D_1 \Delta t \right]$ operation: 
\begin{eqnarray}
\tilde{\bs{r}}_i
&\leftarrow&
\tilde{\bs{r}}_i + \frac{\tilde{\bs{p}}_i}{m_isV^{\frac{2}{3}}} \Delta t~,
\\
\zeta_{i1} 
&\leftarrow&
\frac{1}{4I_1 s} \bs{\pi}'^{\rm T}_i {\stackrel{\leftrightarrow}{\bcal P}}_1 \bs{q}_i~,
\\
\bs{q}_i 
&\leftarrow&
\cos \left( \zeta_{i1} \Delta t \right) \bs{q}_i + 
\sin \left( \zeta_{i1} \Delta t \right) {\stackrel{\leftrightarrow}{\bcal P}}_1 \bs{q}_i~,
\\
\bs{\pi}'_i 
&\leftarrow&
\cos \left( \zeta_{i1} \Delta t \right) \bs{\pi}'_i + 
\sin \left( \zeta_{i1} \Delta t \right) {\stackrel{\leftrightarrow}{\bcal P}}_1 \bs{\pi}'_i~,
\\
P_s 
&\leftarrow&
P_s + 
\left( \sum_{i=1}^N \frac{\tilde{\bs{p}}^2_i}{2m_is^2V^{\frac{2}{3}}} + 
       \sum_{i=1}^N 2I_1 \zeta^2_{i1} 
     - g k_{\rm B}T_0 \log s + H_0 - g k_{\rm B}T_0
\right) \Delta t~,
\\
P_V 
&\leftarrow&
P_V + \sum_{i=1}^N \frac{\tilde{\bs{p}}^2_i}{3m_isV^{\frac{5}{3}}} \Delta t~.
\end{eqnarray}
Step 7. ${\rm exp} \left[ D_2 \Delta t /2 \right]$ operation: 
\begin{eqnarray}
\zeta_{i2} 
&\leftarrow&
\frac{1}{4I_2 s} \bs{\pi}'^{\rm T}_i {\stackrel{\leftrightarrow}{\bcal P}}_2 \bs{q}_i~,
\\
\bs{q}_i 
&\leftarrow&
\cos \left( \zeta_{i2} \frac{\Delta t}{2} \right) \bs{q}_i + 
\sin \left( \zeta_{i2} \frac{\Delta t}{2} \right) {\stackrel{\leftrightarrow}{\bcal P}}_2 \bs{q}_i~,
\\
\bs{\pi}'_i 
&\leftarrow&
\cos \left( \zeta_{i2} \frac{\Delta t}{2} \right) \bs{\pi}'_i + 
\sin \left( \zeta_{i2} \frac{\Delta t}{2} \right) {\stackrel{\leftrightarrow}{\bcal P}}_2 \bs{\pi}'_i~,
\\
P_s 
&\leftarrow&
P_s + 
\left( \sum_{i=1}^N 2I_2 \zeta^2_{i2} \right) 
\frac{\Delta t}{2}~.
\end{eqnarray}
Step 8. ${\rm exp} \left[ D_3 \Delta t /2 \right]$ operation: 
\begin{eqnarray}
\zeta_{i3} 
&\leftarrow&
\frac{1}{4I_3 s} \bs{\pi}'^{\rm T}_i {\stackrel{\leftrightarrow}{\bcal P}}_3 \bs{q}_i~,
\\
\bs{q}_i 
&\leftarrow&
\cos \left( \zeta_{i3} \frac{\Delta t}{2} \right) \bs{q}_i + 
\sin \left( \zeta_{i3} \frac{\Delta t}{2} \right) {\stackrel{\leftrightarrow}{\bcal P}}_3 \bs{q}_i~,
\\
\bs{\pi}'_i 
&\leftarrow&
\cos \left( \zeta_{i3} \frac{\Delta t}{2} \right) \bs{\pi}'_i + 
\sin \left( \zeta_{i3} \frac{\Delta t}{2} \right) {\stackrel{\leftrightarrow}{\bcal P}}_3 \bs{\pi}'_i~,
\\
P_s 
&\leftarrow&
P_s + 
\left( \sum_{i=1}^N 2I_3 \zeta^2_{i3} \right) 
\frac{\Delta t}{2}~.
\end{eqnarray}
Step 9. ${\rm exp} \left[ D_4 \Delta t /2 \right]$ operation: 
\begin{eqnarray}
P_s 
&\leftarrow&
P_s - \frac{P^2_V}{2W} \frac{\Delta t}{2}~,
\\
V 
&\leftarrow&
V + s \frac{P_V}{W} \frac{\Delta t}{2}~.
\end{eqnarray}
Step 10. ${\rm exp} \left[ D_5 \Delta t /2 \right]$ operation: 
\begin{eqnarray}
\tilde{\bs{p}}_i
&\leftarrow&
\tilde{\bs{p}}_i + s V^{\frac{1}{3}} \bs{F}_i \frac{\Delta t}{2}~,
\\
\bs{\pi}'_i
&\leftarrow&
\bs{\pi}'_i + 2s {\stackrel{\leftrightarrow}{\bs S}}(\bs{q}_i) \bs{N}^{(4)}_i \frac{\Delta t}{2}~,
\\
P_s 
&\leftarrow&
P_s - \left[ E(\tilde{\bs{r}}^{\{N\}}, \bs{q}^{\{N\}}, V) 
+ P_0V \right] \frac{\Delta t}{2}~,
\\
P_V 
&\leftarrow&
P_V 
+ s \left( \frac{1}{3V} \sum_{i=1}^N \bs{F}_i \cdot \bs{r}_i - P_0 \right) 
    \frac{\Delta t}{2}~.
\end{eqnarray}
Step 11. ${\rm exp} \left[ D_6 \Delta t /2 \right]$ operation: 
\begin{eqnarray}
s 
&\leftarrow&
s \left( 1+ \frac{P_s}{2Q}\frac{\Delta t}{2} \right)^2~,
\\
P_s 
&\leftarrow&
P_s \bigg/ \left( 1+ \frac{P_s}{2Q}\frac{\Delta t}{2} \right)~.
\end{eqnarray}

\subsection{Symplectic molecular dynamics algorithm for rigid-body molecules 
in the constant temperature, constant normal pressure, and 
constant lateral surface area ensemble 
} 
\label{nose-pcr-and1-symp-rg:sec}
An explicit symplectic MD algorithm 
in the constant temperature, constant normal pressure, and 
constant lateral surface area ensemble is also easily obtained. 
In section \ref{nose-pcr-and-symp-rg:sec} 
the Andersen's constant pressure algorithm was employed 
for all three side lengths of the simulation cell. 
On the other hand, 
one of the side lengths of the simulation cell fluctuates 
in the constant normal pressure and constant lateral surface area ensemble. 
This ensemble is often used for membrane systems \cite{ike04}. 
The Hamiltonian for this ensemble is given by 
\begin{eqnarray}
H_{\rm NPA1-RB} &=& s 
                \left[
                \sum_{i=1}^N \frac{\tilde{p}^2_{xi}}
                                  {2m_is^2L^2} + 
                \sum_{i=1}^N \frac{p'^2_{yi}+p'^2_{zi}}{2m_is^2} + 
                \sum_{i=1}^N \frac{1}{8s^2} 
                       \bs{\pi}'^{\rm T}_i 
                       {\stackrel{\leftrightarrow}{\bs S}}(\bs{q}_i) 
                       {\stackrel{\leftrightarrow}{\bs D}}_i 
                       {\stackrel{\leftrightarrow}{\bs S}}^{\rm T}(\bs{q}_i) 
                       \bs{\pi}'_i 
                \right.
                \nonumber \\
            &+& \left. 
                       E(\tilde{x}^{\{N\}}, y^{\{N\}}, z^{\{N\}}, 
                           \bs{q}^{\{N\}}, L) + 
                \frac{P_s^2}{2Q} + g k_{\rm B}T_0 \log s + 
                \frac{P_L^2}{2W} + P_0AL - H_0 
                \right]~,
\label{hnpa1:eq}
\end{eqnarray}
where 
the variable $P_L$ is the conjugate momenta for 
the side length $L$ of the simulation cell along the $x$-axis. 
The constant $A$ is the lateral surface area on the $yz$-plane. 
Therefore the volume of the simulation cell $V$ is given by $AL$.
Note that $x$ components of $\bs{p}_i$ and $\bs{r}_i$ are scaled by 
$p_{xi} = \tilde{p}_{xi}/sL$ 
and 
$x_i    = L \tilde{x}_i~,$ 
respectively, whereas 
$y$ and $z$ components of $\bs{p}_i$ are scaled by Eq.~(\ref{sc1p:eq}). 
The equations of motion are given by 
\begin{eqnarray}
\dot{x}_i &=& \frac{p_{xi}}{m_i} + \frac{\dot{L}}{L}x_i~, 
              \label{eomnpa11:eq} \\
\dot{y}_i &=& \frac{p_{yi}}{m_i}~, \ \ 
\dot{z}_i  =  \frac{p_{zi}}{m_i}~, \label{eomnpa13:eq} \\
\dot{p}_{xi} &=& F_{xi} 
               - \left( \frac{\dot{s}}{s} + \frac{\dot{L}}{L} \right) 
                 p_{xi}~, \label{eomnpa14:eq} \\
\dot{p}_{yi} &=& F_{yi} - \frac{\dot{s}}{s} p_{yi}~,\ \ 
\dot{p}_{zi}  =  F_{zi} - \frac{\dot{s}}{s} p_{zi}~, \label{eomnpa16:eq} \\
\dot{\bs{q}}_i &=& \frac{1}{2} {\stackrel{\leftrightarrow}{\bs S}}(\bs{q}_i) \bs{\omega}^{(4)}_i~, 
                   \label{eomnpa17:eq} \\
{\stackrel{\leftrightarrow}{{\bs I}_i}} \dot{\bs{\omega}}_i &=& 
\bs{N}_i - \bs{\omega}_i \times \left( {\stackrel{\leftrightarrow}{{\bs I}_i}} \bs{\omega}_i \right)
- \frac{\dot{s}}{s}{\stackrel{\leftrightarrow}{{\bs I}_i}} \bs{\omega}_i~, \label{eomnpa18:eq} \\
\dot{s}        &=& s \frac{P_s}{Q}~, \label{eomnpa19:eq} \\
\dot{P}_s      &=& \sum_{i=1}^N \frac{\bs{p}^2_i}{m_i} 
                 + \sum_{i=1}^N \bs{\omega}^{\rm T}_i {\stackrel{\leftrightarrow}{{\bs I}_i}} \bs{\omega}_i 
                 - g k_{\rm B}T_0~, 
                   \label{eomnpa20:eq} 
\end{eqnarray}
\begin{eqnarray}
\dot{L}        &=& s \frac{P_L}{W}~, \label{eomnpa21:eq} \\
\dot{P}_L      &=& sA\left[ \frac{1}{V} 
                     \left( \sum_{i=1}^N \frac{p^2_{xi}}{m_i} 
                           + \sum_{i=1}^N F_{xi} \cdot x_i 
                     \right)
                 - P_0 \right]~.
                   \label{eomnpa22:eq}
\end{eqnarray}
The Hamiltonian in Eq.~(\ref{hnpa1:eq}) is separated 
into six terms as follows: 
\begin{eqnarray}
H_{\rm NPA1-RB1} &=& s \left[
                     \sum_{i=1}^N \frac{\tilde{p}^2_{xi}}{2m_is^2L^2}
                   + \sum_{i=1}^N \frac{p'^2_{yi}+p'^2_{zi}}{2m_is^2}
                   \right. \nonumber \\
                 & & \ \ \left.
                   + \sum_{i=1}^N \frac{1}{8I_1 s^2} 
                       \left( \bs{\pi}'^{\rm T}_i 
                       {\stackrel{\leftrightarrow}{\bcal P}}_1 \bs{q}_i \right)^2
                   + g k_{\rm B}T_0 \log s - H_0 
                 \right]~, \\
H_{\rm NPA1-RB2} &=& s \sum_{i=1}^N \frac{1}{8I_2 s^2} 
                 \left( \bs{\pi}'^{\rm T}_i {\stackrel{\leftrightarrow}{\bcal P}}_2 \bs{q}_i \right)^2~, \\
H_{\rm NPA1-RB3} &=& s \sum_{i=1}^N \frac{1}{8I_3 s^2} 
                 \left( \bs{\pi}'^{\rm T}_i {\stackrel{\leftrightarrow}{\bcal P}}_3 \bs{q}_i \right)^2~, \\
H_{\rm NPA1-RB4} &=& s \frac{P_L^2}{2W}~, \\
H_{\rm NPA1-RB5} &=& s \left[ 
                  E(\tilde{x}^{\{N\}}, y^{\{N\}}, z^{\{N\}}, \bs{q}^{\{N\}}, L)
                  + P_0AL \right]~, \\
H_{\rm NPA1-RB6} &=& s \frac{P_s^2}{2Q}~.
\end{eqnarray}
In order to obtain the second-order symplectic formula, 
the time propagator ${\rm exp} \left[ D \Delta t \right]$ is again 
decomposed to a product of six time propagators as in Eq.~(\ref{dcmp2:eq}). 
The symplectic time developments are then given by 
\begin{eqnarray}
{\rm exp} \left[ D_1 \Delta t \right]
\tilde{x}_i
&=&
\tilde{x}_i + \frac{\tilde{p}_{xi}}{m_isL^2} \Delta t~,
\\
{\rm exp} \left[ D_1 \Delta t \right]
y_i
&=&
y_i + \frac{p'^2_{yi}}{m_is} \Delta t~,
\\
{\rm exp} \left[ D_1 \Delta t \right]
z_i
&=&
z_i + \frac{p'^2_{zi}}{m_is} \Delta t~,
\\
{\rm exp} \left[ D_1 \Delta t \right]
\bs{q}_i 
&=&
\cos \left( \zeta_{i1} \Delta t \right) \bs{q}_i + 
\sin \left( \zeta_{i1} \Delta t \right) {\stackrel{\leftrightarrow}{\bcal P}}_1 \bs{q}_i~, \ \ 
{\rm where} \
\zeta_{i1} = \frac{1}{4I_1 s} \bs{\pi}'^{\rm T}_i {\stackrel{\leftrightarrow}{\bcal P}}_1 \bs{q}_i~, 
\\
{\rm exp} \left[ D_1 \Delta t \right]
\bs{\pi}'_i 
&=&
\cos \left( \zeta_{i1} \Delta t \right) \bs{\pi}'_i + 
\sin \left( \zeta_{i1} \Delta t \right) {\stackrel{\leftrightarrow}{\bcal P}}_1 \bs{\pi}'_i~,
\\
{\rm exp} \left[ D_1 \Delta t \right]
P_s 
&=&
P_s + 
\left( 
  \sum_{i=1}^N \frac{\tilde{p}^2_{xi}}{2m_is^2L^2}
+ \sum_{i=1}^N \frac{p'^2_{yi}+p'^2_{zi}}{2m_is^2}
\right. \nonumber \\
& & \ \ \ \ \ 
\left.
+ \sum_{i=1}^N 2I_1 \zeta^2_{i1} 
- g k_{\rm B}T_0 \log s + H_0 - g k_{\rm B}T_0 
\right) \Delta t~,
\\
{\rm exp} \left[ D_1 \Delta t \right]
P_L 
&=&
P_L + \sum_{i=1}^N \frac{\tilde{p}^2_{xi}}{m_isL^3} \Delta t~,
\end{eqnarray}
\begin{eqnarray}
{\rm exp} \left[ D_2 \Delta t \right]
\bs{q}_i 
&=&
\cos \left( \zeta_{i2} \Delta t \right) \bs{q}_i + 
\sin \left( \zeta_{i2} \Delta t \right) {\stackrel{\leftrightarrow}{\bcal P}}_2 \bs{q}_i~, \ \ 
{\rm where} \
\zeta_{i2} = \frac{1}{4I_2 s} \bs{\pi}'^{\rm T}_i {\stackrel{\leftrightarrow}{\bcal P}}_2 \bs{q}_i~, 
\\
{\rm exp} \left[ D_2 \Delta t \right]
\bs{\pi}'_i 
&=&
\cos \left( \zeta_{i2} \Delta t \right) \bs{\pi}'_i + 
\sin \left( \zeta_{i2} \Delta t \right) {\stackrel{\leftrightarrow}{\bcal P}}_2 \bs{\pi}'_i~,
\\
{\rm exp} \left[ D_2 \Delta t \right]
P_s 
&=&
P_s + 
\left( \sum_{i=1}^N 2I_2 \zeta^2_{i2} \right) \Delta t~,
\\
{\rm exp} \left[ D_3 \Delta t \right]
\bs{q}_i 
&=&
\cos \left( \zeta_{i3} \Delta t \right) \bs{q}_i + 
\sin \left( \zeta_{i3} \Delta t \right) {\stackrel{\leftrightarrow}{\bcal P}}_3 \bs{q}_i~, \ \ 
{\rm where} \
\zeta_{i3} = \frac{1}{4I_3 s} \bs{\pi}'^{\rm T}_i {\stackrel{\leftrightarrow}{\bcal P}}_3 \bs{q}_i~, 
\\
{\rm exp} \left[ D_3 \Delta t \right]
\bs{\pi}'_i 
&=&
\cos \left( \zeta_{i3} \Delta t \right) \bs{\pi}'_i + 
\sin \left( \zeta_{i3} \Delta t \right) {\stackrel{\leftrightarrow}{\bcal P}}_3 \bs{\pi}'_i~,
\\
{\rm exp} \left[ D_3 \Delta t \right]
P_s 
&=&
P_s + 
\left( \sum_{i=1}^N 2I_3 \zeta^2_{i3} \right) \Delta t~,
\\
{\rm exp} \left[ D_4 \Delta t \right]
P_s 
&=&
P_s - \frac{P^2_L}{2W} \Delta t~,
\\
{\rm exp} \left[ D_4 \Delta t \right]
L 
&=&
L + s \frac{P_L}{W} \Delta t~,
\\
{\rm exp} \left[ D_5 \Delta t \right]
\tilde{p}_{xi}
&=&
\tilde{p}_{xi} + s L F_{xi} \Delta t~,
\\
{\rm exp} \left[ D_5 \Delta t \right]
p'_{yi}
&=&
p'_{yi} + s F_{yi} \Delta t~,
\\
{\rm exp} \left[ D_5 \Delta t \right]
p'_{zi}
&=&
p'_{zi} + s F_{zi} \Delta t~,
\\
{\rm exp} \left[ D_5 \Delta t \right]
\bs{\pi}'_i
&=&
\bs{\pi}'_i + 2s {\stackrel{\leftrightarrow}{\bs S}}(\bs{q}_i) \bs{N}^{(4)}_i \Delta t~,
\\
{\rm exp} \left[ D_5 \Delta t \right]
P_s 
&=&
P_s - \left( E + P_0AL \right) \Delta t~,
\\
{\rm exp} \left[ D_5 \Delta t \right]
P_L 
&=&
P_L 
+ s \left( \frac{1}{L} \sum_{i=1}^N F_{xi} \cdot x_i - P_0A \right) \Delta t~,
\\
{\rm exp} \left[ D_6 \Delta t \right]
s &=&
s \left( 1+ \frac{P_s}{2Q} \Delta t \right)^2~,
\\
{\rm exp} \left[ D_6 \Delta t \right]
P_s &=&
P_s \bigg/ \left( 1+ \frac{P_s}{2Q} \Delta t \right)~.
\end{eqnarray}

\subsection{Symplectic molecular dynamics algorithm for rigid-body molecules 
combined with the Nos\'e-Poincar\'e thermostat and 
the Parrinello-Rahman barostat} 
\label{nose-pcr-pr-symp-rg:sec}
In this subsection we present an explicit symplectic MD algorithm 
for rigid-body molecules in the isothermal-isobaric ensemble 
with simulation-cell deformation. 
The Hamiltonian is given 
by combing the Hamiltonian in Eq.~(\ref{hnprb:eq}) and 
the Parrinello-Rahman barostat \cite{pr80} as follows: 
\begin{eqnarray}
H_{\rm NPPR-RB} &=& s 
                \left[
                \sum_{i=1}^N \frac{1}{2m_is^2} 
                \tilde{\bs{p}}^{\rm T}_i 
                {\stackrel{\leftrightarrow}{\bs G}}^{-1} \tilde{\bs{p}}_i
                + 
                \sum_{i=1}^N \frac{1}{8s^2} 
                       \bs{\pi}'^{\rm T}_i 
                       {\stackrel{\leftrightarrow}{\bs S}}(\bs{q}_i) 
                       {\stackrel{\leftrightarrow}{\bs D}}_i 
                       {\stackrel{\leftrightarrow}{\bs S}}^{\rm T}(\bs{q}_i) 
                       \bs{\pi}'_i 
                       + E(\tilde{\bs{r}}^{\{N\}}, \bs{q}^{\{N\}}, 
                       {\stackrel{\leftrightarrow}{\bs L}}) 
                \right.
                \nonumber \\
            & & \left. + 
                \frac{P_s^2}{2Q} + g k_{\rm B}T_0 \log s + 
                \frac{1}{2W} {\rm Tr} 
                \left( \stackrel{\leftrightarrow}{\bs P}^{\rm T}_{L} \ 
                \stackrel{\leftrightarrow}{\bs P}_{L} \right)
                + P_0V - H_0 
                \right]~,
\label{hnppr:eq}
\end{eqnarray}
where ${\stackrel{\leftrightarrow}{\bs L}}$ is the matrix of cell parameters, 
$\stackrel{\leftrightarrow}{\bs P}_L$ is the conjugate momenta for 
${\stackrel{\leftrightarrow}{\bs L}}$, and 
${\stackrel{\leftrightarrow}{\bs G}}$ is given by 
${\stackrel{\leftrightarrow}{\bs L}}^{\rm T} \ {\stackrel{\leftrightarrow}{\bs L}}$~. 
The scaled momentum $\tilde{\bs{p}}_i$ and 
the scaled coordinate $\tilde{\bs{r}}_i$ 
are related to $\bs{p}_i$ and $\bs{r}_i$ here by 
\begin{eqnarray}
\bs{p}_i &=& \frac{1}{s} \left({\stackrel{\leftrightarrow}{\bs L}}^{\rm T} 
\right)^{-1} \tilde{\bs{p}}_i~, 
\label{sc3p:eq} \\
\bs{r}_i &=& {\stackrel{\leftrightarrow}{\bs L}} \tilde{\bs{r}}_i~. \label{sc3r:eq} 
\end{eqnarray}
The equations of motion are given by 
\begin{eqnarray}
\dot{\bs{r}}_i &=& \frac{\bs{p}_i}{m_i} 
                 + \dot{\stackrel{\leftrightarrow}{\bs L}}{\stackrel{\leftrightarrow}{\bs L}}^{-1}\bs{r}_i~, 
                 \label{eomnppr1:eq} \\
\dot{\bs{p}}_i &=& \bs{F}_i 
                 - \frac{\dot{s}}{s} \bs{p}_i
                 - \left( \dot{\stackrel{\leftrightarrow}{\bs L}}{\stackrel{\leftrightarrow}{\bs L}}^{-1} \right)^{\rm T} \bs{p}_i~, 
                 \label{eomnppr2:eq} \\
\dot{\bs{q}}_i &=& \frac{1}{2} {\stackrel{\leftrightarrow}{\bs S}}(\bs{q}_i) \bs{\omega}^{(4)}_i~, 
                   \label{eomnppr3:eq} \\
{\stackrel{\leftrightarrow}{{\bs I}_i}} \dot{\bs{\omega}}_i &=& 
\bs{N}_i - 
\bs{\omega}_i \times 
\left( {\stackrel{\leftrightarrow}{{\bs I}_i}} \bs{\omega}_i \right)
- \frac{\dot{s}}{s}{\stackrel{\leftrightarrow}{{\bs I}_i}} \bs{\omega}_i~, 
\label{eomnppr4:eq} \\
\dot{s}        &=& s \frac{P_s}{Q}~, \label{eomnppr5:eq} \\
\dot{P}_s      &=& \sum_{i=1}^N \frac{\bs{p}^2_i}{m_i} 
                 + \sum_{i=1}^N \bs{\omega}^{\rm T}_i 
                 {\stackrel{\leftrightarrow}{{\bs I}_i}} \bs{\omega}_i 
                 - g k_{\rm B}T_0~, 
                   \label{eomnppr6:eq} \\
\dot{\stackrel{\leftrightarrow}{\bs L}}    
&=& \frac{s}{W}\stackrel{\leftrightarrow}{\bs P}_L~, \label{eomnppr7:eq} \\
\dot{\stackrel{\leftrightarrow}{\bs P}}_L  &=& s \left[ \frac{1}{V} 
                     \left( \sum_{i=1}^N \frac{1}{m_i} 
                            \bs{p}_i \bs{p}^{\rm T}_i
                          + \sum_{i=1}^N \bs{F}_i \bs{r}^{\rm T}_i 
                     \right)
                 - P_0 \stackrel{\leftrightarrow}{\bs 1}\right] 
                   \stackrel{\leftrightarrow}{\bs \sigma}~,
                   \label{eomnppr8:eq}
\end{eqnarray}
where $\stackrel{\leftrightarrow}{\bs \sigma}$ 
is related to $\stackrel{\leftrightarrow}{\bs L}$ 
by $\stackrel{\leftrightarrow}{\bs \sigma}^{\rm T} = V{\stackrel{\leftrightarrow}{\bs L}}^{-1}$ and 
$\stackrel{\leftrightarrow}{\bs 1}$ is the identity matrix. 
Note that $\bs{p}_i \bs{p}^{\rm T}_i$ and $\bs{F}_i \bs{r}^{\rm T}_i$ 
are dyadic tensors, whose $(\alpha,\beta)$ elements ($\alpha,\beta=x,y,z$) 
are $p_{\alpha i} p_{\beta i}$ and $F_{\alpha i} r_{\beta i}$, respectively. 
The Hamiltonian in Eq.~(\ref{hnppr:eq}) is also separated 
into six terms as follows: 
\begin{eqnarray}
H_{\rm NPPR-RB1} &=& s \left[
                     \sum_{i=1}^N \frac{1}{2m_is^2} 
                     \tilde{\bs{p}}^{\rm T}_i 
                     {\stackrel{\leftrightarrow}{\bs G}}^{-1} \tilde{\bs{p}}_i
                   + \sum_{i=1}^N \frac{1}{8I_1 s^2} 
                     \left( \bs{\pi}'^{\rm T}_i 
                     {\stackrel{\leftrightarrow}{\bcal P}}_1 \bs{q}_i \right)^2
                   + g k_{\rm B}T_0 \log s - H_0 
                 \right]~, \\
H_{\rm NPPR-RB2} &=& s \sum_{i=1}^N \frac{1}{8I_2 s^2} 
                 \left( \bs{\pi}'^{\rm T}_i 
                 {\stackrel{\leftrightarrow}{\bcal P}}_2 \bs{q}_i \right)^2~, \\
H_{\rm NPPR-RB3} &=& s \sum_{i=1}^N \frac{1}{8I_3 s^2} 
                 \left( \bs{\pi}'^{\rm T}_i 
                 {\stackrel{\leftrightarrow}{\bcal P}}_3 \bs{q}_i \right)^2~, \\
H_{\rm NPPR-RB4} &=& 
                 \frac{s}{2W} {\rm Tr} 
                 \left( \stackrel{\leftrightarrow}{\bs P}^{\rm T}_{L} \ 
                 \stackrel{\leftrightarrow}{\bs P}_{L} \right)~, \\
H_{\rm NPPR-RB5} &=& s \left[ 
                    E(\tilde{\bs{r}}^{\{N\}}, \bs{q}^{\{N\}}, 
                     {\stackrel{\leftrightarrow}{\bs L}}) 
                  + P_0V \right]~, \\
H_{\rm NPPR-RB6} &=& s \frac{P_s^2}{2Q}~.
\end{eqnarray}
The symplectic time developments are given using 
the decomposition of ${\rm exp} \left[ D \Delta t \right]$ 
in Eq.~(\ref{dcmp2:eq}) by 
\begin{eqnarray}
{\rm exp} \left[ D_1 \Delta t \right]
\tilde{\bs{r}}_i
&=&
\tilde{\bs{r}}_i + \frac{\Delta t}{m_i s} {\stackrel{\leftrightarrow}{\bs G}}^{-1} \tilde{\bs{p}}_i~,
\\
{\rm exp} \left[ D_1 \Delta t \right]
\bs{q}_i 
&=&
\cos \left( \zeta_{i1} \Delta t \right) \bs{q}_i + 
\sin \left( \zeta_{i1} \Delta t \right) {\stackrel{\leftrightarrow}{\bcal P}}_1 \bs{q}_i~, \ \ 
{\rm where} \
\zeta_{i1} = \frac{1}{4I_1 s} \bs{\pi}'^{\rm T}_i {\stackrel{\leftrightarrow}{\bcal P}}_1 \bs{q}_i~, 
\\
{\rm exp} \left[ D_1 \Delta t \right]
\bs{\pi}'_i 
&=&
\cos \left( \zeta_{i1} \Delta t \right) \bs{\pi}'_i + 
\sin \left( \zeta_{i1} \Delta t \right) {\stackrel{\leftrightarrow}{\bcal P}}_1 \bs{\pi}'_i~,
\\
{\rm exp} \left[ D_1 \Delta t \right]
P_s 
&=&
P_s + 
\left( \sum_{i=1}^N \frac{1}{2m_is^2} 
       \tilde{\bs{p}}^{\rm T}_i 
       {\stackrel{\leftrightarrow}{\bs G}}^{-1} \tilde{\bs{p}}_i 
\right. \nonumber \\
& & \ \ \ \ \ + 
\left.
       \sum_{i=1}^N 2I_1 \zeta^2_{i1} 
     - g k_{\rm B}T_0 \log s + H_0 - g k_{\rm B}T_0
\right) \Delta t~,
\\
{\rm exp} \left[ D_1 \Delta t \right]
\stackrel{\leftrightarrow}{\bs P}_{L}
&=&
\stackrel{\leftrightarrow}{\bs P}_{L} + \frac{\Delta t}{V}
\left\{ \sum_{i=1}^N \frac{1}{m_i s} 
        \left[ \left( \stackrel{\leftrightarrow}{\bs L}^{\rm T} \right)^{-1} 
                \tilde{\bs{p}}_i \right]
        \left[ \left( \stackrel{\leftrightarrow}{\bs L}^{\rm T} \right)^{-1} 
                \tilde{\bs{p}}_i \right]^{\rm T}
\right\} \stackrel{\leftrightarrow}{\bs \sigma}~,
\\
{\rm exp} \left[ D_2 \Delta t \right]
\bs{q}_i 
&=&
\cos \left( \zeta_{i2} \Delta t \right) \bs{q}_i + 
\sin \left( \zeta_{i2} \Delta t \right) {\stackrel{\leftrightarrow}{\bcal P}}_2 \bs{q}_i~, \ \ 
{\rm where} \
\zeta_{i2} = \frac{1}{4I_2 s} \bs{\pi}'^{\rm T}_i {\stackrel{\leftrightarrow}{\bcal P}}_2 \bs{q}_i~, 
\\
{\rm exp} \left[ D_2 \Delta t \right]
\bs{\pi}'_i 
&=&
\cos \left( \zeta_{i2} \Delta t \right) \bs{\pi}'_i + 
\sin \left( \zeta_{i2} \Delta t \right) {\stackrel{\leftrightarrow}{\bcal P}}_2 \bs{\pi}'_i~,
\\
{\rm exp} \left[ D_2 \Delta t \right]
P_s 
&=&
P_s + 
\left( \sum_{i=1}^N 2I_2 \zeta^2_{i2} \right) 
\Delta t~,
\\
{\rm exp} \left[ D_3 \Delta t \right]
\bs{q}_i 
&=&
\cos \left( \zeta_{i3} \Delta t \right) \bs{q}_i + 
\sin \left( \zeta_{i3} \Delta t \right) {\stackrel{\leftrightarrow}{\bcal P}}_3 \bs{q}_i~, \ \ 
{\rm where} \
\zeta_{i3} = \frac{1}{4I_3 s} \bs{\pi}'^{\rm T}_i {\stackrel{\leftrightarrow}{\bcal P}}_3 \bs{q}_i~, 
\\
{\rm exp} \left[ D_3 \Delta t \right]
\bs{\pi}'_i 
&=&
\cos \left( \zeta_{i3} \Delta t \right) \bs{\pi}'_i + 
\sin \left( \zeta_{i3} \Delta t \right) {\stackrel{\leftrightarrow}{\bcal P}}_3 \bs{\pi}'_i~,
\\
{\rm exp} \left[ D_3 \Delta t \right]
P_s 
&=&
P_s + 
\left( \sum_{i=1}^N 2I_3 \zeta^2_{i3} \right) 
\Delta t~,
\end{eqnarray}
\begin{eqnarray}
{\rm exp} \left[ D_4 \Delta t \right]
P_s 
&=&
P_s - \frac{\Delta t}{2W} 
{\rm Tr} \left( \stackrel{\leftrightarrow}{\bs P}^{\rm T}_{L} \ 
                \stackrel{\leftrightarrow}{\bs P}_{L} \right)~,
\\
{\rm exp} \left[ D_4 \Delta t \right]
{\stackrel{\leftrightarrow}{\bs L}}
&=&
{\stackrel{\leftrightarrow}{\bs L}} + \frac{s \Delta t}{W} \stackrel{\leftrightarrow}{\bs P}_{L}~,
\\
{\rm exp} \left[ D_5 \Delta t \right]
\tilde{\bs{p}}_i
&=&
\tilde{\bs{p}}_i + s {\stackrel{\leftrightarrow}{\bs L}}^{\rm T} \bs{F}_i \Delta t~,
\\
{\rm exp} \left[ D_5 \Delta t \right]
\bs{\pi}'_i
&=&
\bs{\pi}'_i + 2s {\stackrel{\leftrightarrow}{\bs S}}(\bs{q}_i) \bs{N}^{(4)}_i \Delta t~,
\\
{\rm exp} \left[ D_5 \Delta t \right]
P_s 
&=&
P_s - \left[ E(\tilde{\bs{r}}^{\{N\}}, \bs{q}^{\{N\}}, {\stackrel{\leftrightarrow}{\bs L}}) 
+ P_0V \right] \Delta t~,
\\
{\rm exp} \left[ D_5 \Delta t \right]
\stackrel{\leftrightarrow}{\bs P}_{L}
&=&
\stackrel{\leftrightarrow}{\bs P}_{L}
+ s \Delta t 
\left( \frac{1}{V} \sum_{i=1}^N \bs{F}_i \bs{r}^{\rm T}_i 
- P_0 \stackrel{\leftrightarrow}{\bs 1} \right)
\stackrel{\leftrightarrow}{\bs \sigma}~,
\\
{\rm exp} \left[ D_6 \Delta t \right]
s 
&=&
s \left( 1+ \frac{P_s}{2Q}\Delta t \right)^2~,
\\
{\rm exp} \left[ D_6 \Delta t \right]
P_s 
&=&
P_s \bigg/ \left( 1+ \frac{P_s}{2Q}\Delta t \right)~.
\end{eqnarray}

\subsection{Symplectic condition and time reversibility} 
\label{symp-time:sec}
In this section we discuss the symplectic condition and 
the time reversibility \cite{gold-symp}. 
Let us suppose a time-independent canonical transformation from
\begin{equation}
\bs{\Gamma} = 
\left(
\begin{array}{c}
\bs{Q} \\ 
\bs{P} \\ 
\end{array} 
\right)
\end{equation}
to 
\begin{equation}
\bs{\Gamma}' = 
\left(
\begin{array}{c}
\bs{Q}'(\bs{Q},\bs{P}) \\ 
\bs{P}'(\bs{Q},\bs{P}) \\ 
\end{array} 
\right)~,
\end{equation}
where $\bs{Q}$ and $\bs{P}$ are the generalized coordinate and 
the generalized momentum, respectively. 
The canonical equation of $\bs{\Gamma}$ is given by 
\begin{equation}
\dot{\bs{\Gamma}} = \bs{J} \frac{\partial H}{\partial \bs{\Gamma}}~,
\end{equation}
where
\begin{equation}
\bs{J} = 
\left(
\begin{array}{cc}
 \bs{0} & \bs{1} \\ 
-\bs{1} & \bs{0} \\ 
\end{array} 
\right)~.
\end{equation}
Because $\bs{\Gamma}'$ is given by the canonical transformation 
from $\bs{\Gamma}$, 
the canonical equation of $\bs{\Gamma}'$ is also given by 
\begin{equation}
\dot{\bs{\Gamma}'} = \bs{J} \frac{\partial H}{\partial \bs{\Gamma}'}~.
\label{eta1:eq}
\end{equation}

The time derivative of $\bs{\Gamma}'(\bs{\Gamma})$ is derived in another way 
by the chain rule: 
\begin{equation}
\dot{\bs{\Gamma}'} = 
\frac{\partial \bs{\Gamma}'}{\partial \bs{\Gamma}}\dot{\bs{\Gamma}} = 
\bs{M} \dot{\bs{\Gamma}} = 
\bs{M} \bs{J} \frac{\partial H}{\partial \bs{\Gamma}} = 
\bs{M} \bs{J} \bs{M}^{\rm T} \frac{\partial H}{\partial \bs{\Gamma}'}~,
\label{eta2:eq}
\end{equation}
where $\bs{M}$ is the Jacobian matrix for the canonical transformation 
from $\bs{\Gamma}$ to $\bs{\Gamma}'$ and its $(i,j)$ element is given by 
\begin{equation}
M_{ij} = \frac{\partial {\it \Gamma}'_i}{\partial {\it \Gamma}_j}~.
\end{equation}
Comparing Eq.~(\ref{eta1:eq}) and Eq.~(\ref{eta2:eq}), 
we obtain the symplectic condition: 
\begin{equation}
\bs{M} \bs{J} \bs{M}^{\rm T} = \bs{J}~.
\label{sym-con:eq}
\end{equation}
In general, 
the generalized coordinates and momenta 
obtained by a Hamiltonian dynamics fulfills 
the symplectic condition in Eq.~(\ref{sym-con:eq}). 

Each factor in the decompositions in Eqs.~(\ref{dcmp1:eq}) 
and (\ref{dcmp2:eq}) is a time propagator based on 
the corresponding Hamiltonian. 
For example, ${\rm exp} \left[ D_1 \Delta t \right]$ in Eqs.~(\ref{dcmp1:eq}) 
is a time propagator by the Hamiltonian of $H_{\rm NP-RB1}$. 
Therefore, the time developments by the decompositions 
in Eqs.~(\ref{dcmp1:eq}) and (\ref{dcmp2:eq})
fulfill the symplectic condition. 
All variables in Eqs.~(\ref{intss-1:eq})-(\ref{intss-f:eq}) are 
canonical variables such as 
$\bs{r}_i$, $\bs{p}'_i$, $\bs{q}_i$, $\bs{\pi}'_i$, $s$, and $P_s$. 
Besides, the time propagator here is decomposed so that 
the MD algorithm will be time reversible, namely, 
${\rm exp} \left[-D \Delta t \right]{\rm exp} \left[ D \Delta t \right]=1$ 
holds in Eqs.~(\ref{dcmp1:eq}) and (\ref{dcmp2:eq}).

Employing the symplectic MD algorithm, 
there is a conserved quantity 
which is close to the Hamiltonian \cite{yoshida90}. 
It means that the long-time deviation of the Hamiltonian is suppressed. 
Therefore, we can perform a MD simulation more stably 
than by conventional nonsymplectic algorithms.

From the symplectic condition in Eq.~(\ref{sym-con:eq}), 
the Jacobian determinant is calculated as one: 
\begin{equation}
{\rm det} \bs{M} = 1~.
\label{jcb1:eq}
\end{equation}
It means that the phase-space volume is conserved during the simulation. 
Note that the phase-space-volume conservation is 
a necessary condition of the symplectic condition 
and not a sufficient condition. 
The condition that the Jacobian determinant is one does not 
always mean symplectic. 
Even if the Jacobian determinant is one, 
there is not always a conserved quantity which is close to the Hamiltonian. 
In other words, 
there are nonsymplectic MD algorithms 
which are phase-space volume conserving and time reversible. 
The time propagators in these nonsymplectic algorithms 
are not based on Hamiltonian and 
the variables are not canonical variables. 
That is, the symplectic condition in Eq.~(\ref{sym-con:eq}) is not fulfilled. 
Therefore, there is not a conserved quantity 
which is close to the Hamiltonian. 
It means that the value of the Hamiltonian deviates gradually
from its initial value in a long-time simulation. 
In the next section we compare our symplectic algorithm 
with the nonsymplectic time-reversible algorithms.

%
 \section{Comparisons with Nonsymplectic Time-Reversible Algorithms} 
 \label{comp:sec}
%
In this section we explain three nonsymplectic 
algorithms in the canonical ensemble, 
which are time reversible. 
We then apply our symplectic algorithm and 
these nonsymplectic algorithms 
to a rigid-body water model and compare them numerically. 

\subsection{Molecular dynamics algorithm based on 
the Nos\'e-Poincar\'e Thermostat 
and the Nonsymplectic Rigid-Body Algorithm} 
\label{nose-pcr-nonsymp-rg:sec}
Instead of the symplectic rigid-body 
MD algorithm by Miller $et$ $al$. \cite{mill02}, 
we here combine the nonsymplectic rigid-body 
MD algorithm by Matubayasi and Nakahara \cite{matu99} 
with the Nos\'e-Poincar\'e thermostat \cite{bond99,nose01}. 
In this algorithm, angular velocity $\bs{\omega}'_i \equiv s \bs{\omega}_i$ 
instead of $\bs{\pi}'_i$ is employed, that is, 
the variables here are 
$\bs{r}_i$, $\bs{p}'_i$, $\bs{q}_i$, $\bs{\omega}'_i$, $s$, and $P_s$. 

The time propagator {\rm exp} $\left[ D \Delta t \right]$ 
is decomposed by 
\begin{eqnarray}
{\rm exp} \left[ D \Delta t \right] 
&=&
{\rm exp} \left[ D_5 \frac{\Delta t}{2} \right]
{\rm exp} \left[ D_4 \frac{\Delta t}{2} \right]
{\rm exp} \left[ D_3 \frac{\Delta t}{2} \right]
{\rm exp} \left[ D_2 \frac{\Delta t}{2} \right]
{\rm exp} \left[ D_1 \Delta t \right] \nonumber \\
&\times&
{\rm exp} \left[ D_2 \frac{\Delta t}{2} \right]
{\rm exp} \left[ D_3 \frac{\Delta t}{2} \right]
{\rm exp} \left[ D_4 \frac{\Delta t}{2} \right]
{\rm exp} \left[ D_5 \frac{\Delta t}{2} \right] \nonumber \\
&+& O \left( \left( \Delta t \right)^3 \right)~.
\label{dcmp3:eq}
\end{eqnarray}
where each time propagator is given by 
\begin{eqnarray}
D_1
&=&
\sum_{i=1}^N \frac{\bs{p}'_i}{m_is} \cdot \frac{\partial }{\partial \bs{r}_i} 
+
\sum_{i=1}^N \frac{1}{2s} 
       \left( {\stackrel{\leftrightarrow}{\bs S}}(\bs{q}_i) \bs{\omega}'^{(4)}_i \right) \cdot 
       \frac{\partial }{\partial \bs{q}_i} \nonumber \\
&+&
\left[ \sum_{i=1}^N \frac{\bs{p}'^2_i}{2m_is^2} + 
       \sum_{i=1}^N \frac{1}{2s^2} 
              \bs{\omega}'^{(4) \rm T}_i {\stackrel{\leftrightarrow}{\bs D}}_i \bs{\omega}'^{(4)}_i
     - g k_{\rm B}T_0 \log s + H_0 + g k_{\rm B}T_0
\right]
\frac{\partial }{\partial P_s}~,
\label{intsn-1:eq}
\\
D_2
&=&
\sum_{i=1}^N \frac{I_{iy} - I_{iz}}{I_{ix} s} \omega'_{iy} \omega'_{iz} 
       \frac{\partial }{\partial \omega'_{ix}} 
+
\sum_{i=1}^N \frac{I_{iz} - I_{iy}}{I_{iz} s} \omega'_{ix} \omega'_{iy} 
       \frac{\partial }{\partial \omega'_{iz}}~, \\
D_3
&=&
\sum_{i=1}^N \frac{I_{iz} - I_{ix}}{I_{iy} s} \omega'_{iz} \omega'_{ix} 
       \frac{\partial }{\partial \omega'_{iy}} 
+
\sum_{i=1}^N \frac{I_{ix} - I_{iz}}{I_{iz} s} \omega'_{ix} \omega'_{iy} 
       \frac{\partial }{\partial \omega'_{iz}}~, \\
D_4
&=&
 \sum_{i=1}^N s \bs{F}_i \cdot \frac{\partial }{\partial \bs{p}'_i} 
+\sum_{i=1}^N s 
 \left( \bs{I}^{-1}_i\bs{N}_i \right) \cdot 
 \frac{\partial }{\partial \bs{\omega}'_i} 
- \sum_{i=1}^N E(\bs{r}^{\{N\}}, \bs{q}^{\{N\}}) 
               \frac{\partial }{\partial P_s}~, \\
D_5
&=&
\frac{s P_s}{Q} \frac{\partial }{\partial s} - 
\frac{P^2_s}{2Q} \frac{\partial }{\partial P_s}~. 
\label{intsn-f:eq}
\end{eqnarray}

\subsection{Molecular dynamics algorithm based on 
the Nos\'e-Hoover Thermostat 
and the Symplectic Rigid-Body Algorithm} 
\label{nose-hvr-symp-rg:sec}
We here combine the symplectic rigid-body MD algorithm \cite{mill02} 
with the Nos\'e-Hoover thermostat \cite{nose84,nosejcp84,hoo85,martyna96} 
(the latter is nonsymplectic). 
This combination has been employed in Ref.~[\onlinecite{ike04}]. 
Instead of $s$ and $P_s$, 
$\eta=\log s$ and $\xi=P_s/Q$ are used for the thermostat, 
that is, 
the variables employed here are 
$\bs{r}_i$, $\bs{p}_i$, $\bs{q}_i$, $\bs{\pi}_i$, $\eta$, and $\xi$. 

The time propagator {\rm exp} $\left[ D \Delta t \right]$ 
is decomposed by \cite{martyna96} 
\begin{eqnarray}
{\rm exp} \left[ D \Delta t \right] 
&=&
{\rm exp} \left[ D_6 \frac{\Delta t}{4} \right]
{\rm exp} \left[ D_5 \frac{\Delta t}{2} \right]
{\rm exp} \left[ D_6 \frac{\Delta t}{4} \right] 
{\rm exp} \left[ D_4 \frac{\Delta t}{2} \right]
\nonumber \\
&\times&
{\rm exp} \left[ D_3 \frac{\Delta t}{2} \right]
{\rm exp} \left[ D_2 \frac{\Delta t}{2} \right]
{\rm exp} \left[ D_1 \Delta t \right] 
{\rm exp} \left[ D_2 \frac{\Delta t}{2} \right]
{\rm exp} \left[ D_3 \frac{\Delta t}{2} \right]
\nonumber \\
&\times&
{\rm exp} \left[ D_4 \frac{\Delta t}{2} \right] 
{\rm exp} \left[ D_6 \frac{\Delta t}{4} \right]
{\rm exp} \left[ D_5 \frac{\Delta t}{2} \right]
{\rm exp} \left[ D_6 \frac{\Delta t}{4} \right]
\nonumber \\
&+& O \left( \left( \Delta t \right)^3 \right)~, 
\label{dcmp4:eq}
\end{eqnarray}
where each time propagator is given by \cite{ike04} 
\begin{eqnarray}
D_1
&=&
\sum_{i=1}^N \frac{\bs{p}_i}{m_i} \cdot \frac{\partial }{\partial \bs{r}_i} 
\nonumber \\
&+&
\sum_{i=1}^N \frac{1}{4I_1} 
       \left( \bs{\pi}^{\rm T}_i {\stackrel{\leftrightarrow}{\bcal P}}_1 \bs{q}_i \right) 
       \left( {\stackrel{\leftrightarrow}{\bcal P}}_1 \bs{q}_i \right) \cdot 
       \frac{\partial }{\partial \bs{q}_i} 
+
\sum_{i=1}^N \frac{1}{4I_1} 
       \left( \bs{\pi}^{\rm T}_i {\stackrel{\leftrightarrow}{\bcal P}}_1 \bs{q}_i \right) 
       \left( {\stackrel{\leftrightarrow}{\bcal P}}_1 \bs{\pi}_i \right) \cdot 
       \frac{\partial }{\partial \bs{\pi}_i}~,
\label{intns-1:eq}
\\
D_2
&=&
\sum_{i=1}^N \frac{1}{4I_2} 
       \left( \bs{\pi}^{\rm T}_i {\stackrel{\leftrightarrow}{\bcal P}}_2 \bs{q}_i \right) 
       \left( {\stackrel{\leftrightarrow}{\bcal P}}_2 \bs{q}_i \right) \cdot 
       \frac{\partial }{\partial \bs{q}_i} 
+
\sum_{i=1}^N \frac{1}{4I_2} 
       \left( \bs{\pi}^{\rm T}_i {\stackrel{\leftrightarrow}{\bcal P}}_2 \bs{q}_i \right) 
       \left( {\stackrel{\leftrightarrow}{\bcal P}}_2 \bs{\pi}_i \right) \cdot 
       \frac{\partial }{\partial \bs{\pi}_i}~,
\\
D_3
&=&
\sum_{i=1}^N \frac{1}{4I_3} 
       \left( \bs{\pi}^{\rm T}_i {\stackrel{\leftrightarrow}{\bcal P}}_3 \bs{q}_i \right) 
       \left( {\stackrel{\leftrightarrow}{\bcal P}}_3 \bs{q}_i \right) \cdot 
       \frac{\partial }{\partial \bs{q}_i} 
+
\sum_{i=1}^N \frac{1}{4I_3} 
       \left( \bs{\pi}^{\rm T}_i {\stackrel{\leftrightarrow}{\bcal P}}_3 \bs{q}_i \right) 
       \left( {\stackrel{\leftrightarrow}{\bcal P}}_3 \bs{\pi}_i \right) \cdot 
       \frac{\partial }{\partial \bs{\pi}_i}~,
\\
D_4
&=&
\sum_{i=1}^N \bs{F}_i \cdot \frac{\partial }{\partial \bs{p}_i} 
+
\sum_{i=1}^N 2 \left( {\stackrel{\leftrightarrow}{\bs S}}(\bs{q}_i) \bs{N}^{(4)}_i \right)
\cdot \frac{\partial }{\partial \bs{\pi}_i}~,
\\
D_5
&=&
- \xi \sum_{i=1}^N \bs{p}_i \cdot \frac{\partial }{\partial \bs{p}_i} 
- \xi \sum_{i=1}^N \bs{\pi}_i \cdot \frac{\partial }{\partial \bs{\pi}_i} 
+ \xi \frac{\partial }{\partial \eta}~,
\\
D_6
&=&
\frac{1}{Q}
\left( \sum_{i=1}^N \frac{\bs{p}^2_i}{m_i}
     + \sum_{i=1}^N \frac{1}{4} \bs{\pi}^{\rm T}_i {\stackrel{\leftrightarrow}{\bs S}}(\bs{q}_i) {\stackrel{\leftrightarrow}{\bs D}}_i 
                          {\stackrel{\leftrightarrow}{\bs S}}^{\rm T}(\bs{q}_i) \bs{\pi}_i 
     - g k_{\rm B} T_0
\right)
\frac{\partial }{\partial \xi}~.
\label{intns-f:eq}
\end{eqnarray}
We remark that we can also make another second-order integrator 
by the decomposition in Eq.~(\ref{dcmp2:eq}) 
instead of Eq.~(\ref{dcmp4:eq}). 
However, the original time reversible algorithm 
for the Nos\'e-Hoover thermostat decomposed the time propagator 
as in Eq.~(\ref{dcmp4:eq}) \cite{martyna96}, 
thus we used this decomposition.

\subsection{Molecular dynamics algorithm based on 
the Nos\'e-Hoover Thermostat 
and the Nonsymplectic Rigid-Body Algorithm} 
\label{nose-hvr-nonsymp-rg:sec}
We can also make a nonsymplectic algorithm 
by the rigid-body algorithm by Matubayasi and Nakahara \cite{matu99} 
and the Nos\'e-Hoover thermostat \cite{nose84,nosejcp84,hoo85,martyna96}. 
In this algorithm 
the following variables are developed with time: 
$\bs{r}_i$, $\bs{p}_i$, $\bs{q}_i$, $\bs{\omega}_i$, $\eta$, and $\xi$. 

The time propagator {\rm exp} $\left[ D \Delta t \right]$ 
is decomposed as in Eq.~(\ref{dcmp4:eq}). 
Each decomposed time propagator is given by 
\begin{eqnarray}
D_1
&=&
\sum_{i=1}^N \frac{\bs{p}_i}{m_i} \cdot \frac{\partial }{\partial \bs{r}_i} 
+
\sum_{i=1}^N \frac{1}{2} \left( {\stackrel{\leftrightarrow}{\bs S}}(\bs{q}_i) \bs{\omega}^{(4)}_i \right) 
      \cdot \frac{\partial }{\partial \bs{q}_i}~,
\label{intnn-1:eq}
\\
D_2
&=&
\sum_{i=1}^N \frac{I_{iy} - I_{iz}}{I_{ix}} \omega_{iy} \omega_{iz} 
       \frac{\partial }{\partial \omega_{ix}} 
+
\sum_{i=1}^N \frac{I_{iz} - I_{iy}}{I_{iz}} \omega_{ix} \omega_{iy} 
       \frac{\partial }{\partial \omega_{iz}}~, \\
D_3
&=&
\sum_{i=1}^N \frac{I_{iz} - I_{ix}}{I_{iy}} \omega_{iz} \omega_{ix} 
       \frac{\partial }{\partial \omega_{iy}} 
+
\sum_{i=1}^N \frac{I_{ix} - I_{iz}}{I_{iz}} \omega_{ix} \omega_{iy} 
       \frac{\partial }{\partial \omega_{iz}}~, \\
D_4
&=&
  \sum_{i=1}^N \bs{F}_i \cdot \frac{\partial }{\partial \bs{p}_i}
+ \sum_{i=1}^N \left( \bs{I}^{-1}_i\bs{N}_i \right) 
  \cdot \frac{\partial }{\partial \bs{\omega}_i}~,
\\
D_5
&=&
- \xi \sum_{i=1}^N \bs{p}_i \cdot \frac{\partial }{\partial \bs{p}_i} 
- \xi \sum_{i=1}^N \bs{\omega}_i \cdot 
      \frac{\partial }{\partial \bs{\omega}_i} 
+ \xi \frac{\partial }{\partial \eta}~,
\\
D_6
&=&
\frac{1}{Q}
\left( \sum_{i=1}^N \frac{\bs{p}^2_i}{m_i}
     + \sum_{i=1}^N \bs{\omega}^{(4) \rm T}_i {\stackrel{\leftrightarrow}{\bs D}}_i \bs{\omega}^{(4)}_i 
     - g k_{\rm B} T_0
\right)
\frac{\partial }{\partial \xi}~.
\label{intnn-f:eq}
\end{eqnarray}

\subsection{Numerical comparisons: application to a pure water system } 
\label{appl:sec}
We applied the symplectic and nonsymplectic MD algorithms 
to a rigid-body model of water in the canonical ensemble. 
We employed the TIP3P rigid-body model for the water molecules \cite{jorg83}. 
We used 80 water molecules in a cubic unit cell 
with periodic boundary conditions. 
The temperature was set at 300~K and 
the mass density was set to 0.997~g/cm$^3$.
The electrostatic potential was calculated by the Ewald method. 
We calculated the van der Waals interaction, 
which is given by the Lennerd-Jones term, 
of all pairs of the molecules within the minimum image convention 
instead of introducing the spherical potential cutoff. 
We tested the time steps of $\Delta t$ = 2~fs, 3~fs, 4~fs, and 5~fs. 
We performed the MD simulations for 1.5~ns in all cases of $\Delta t$. 
We employed 
Eqs.(\ref{intss-1:eq}) - (\ref{intss-f:eq}) 
for the Nos\'e-Poincar\'e thermostat and 
symplectic rigid-body MD simulations, 
Eqs.(\ref{intsn-1:eq}) - (\ref{intsn-f:eq}) 
for the Nos\'e-Poincar\'e thermostat and 
nonsymplectic rigid-body MD simulations, 
Eqs.(\ref{intns-1:eq}) - (\ref{intns-f:eq}) 
for the Nos\'e-Hoover thermostat and 
symplectic rigid-body MD simulations, and 
the time development in Eqs.(\ref{intnn-1:eq}) - (\ref{intnn-f:eq}) 
for the Nos\'e-Hoover thermostat and 
nonsymplectic rigid-body MD simulations. 
The same initial conditions were used for all algorithms and time steps.

\begin{figure}
\includegraphics[angle=-90,width=15cm,keepaspectratio]{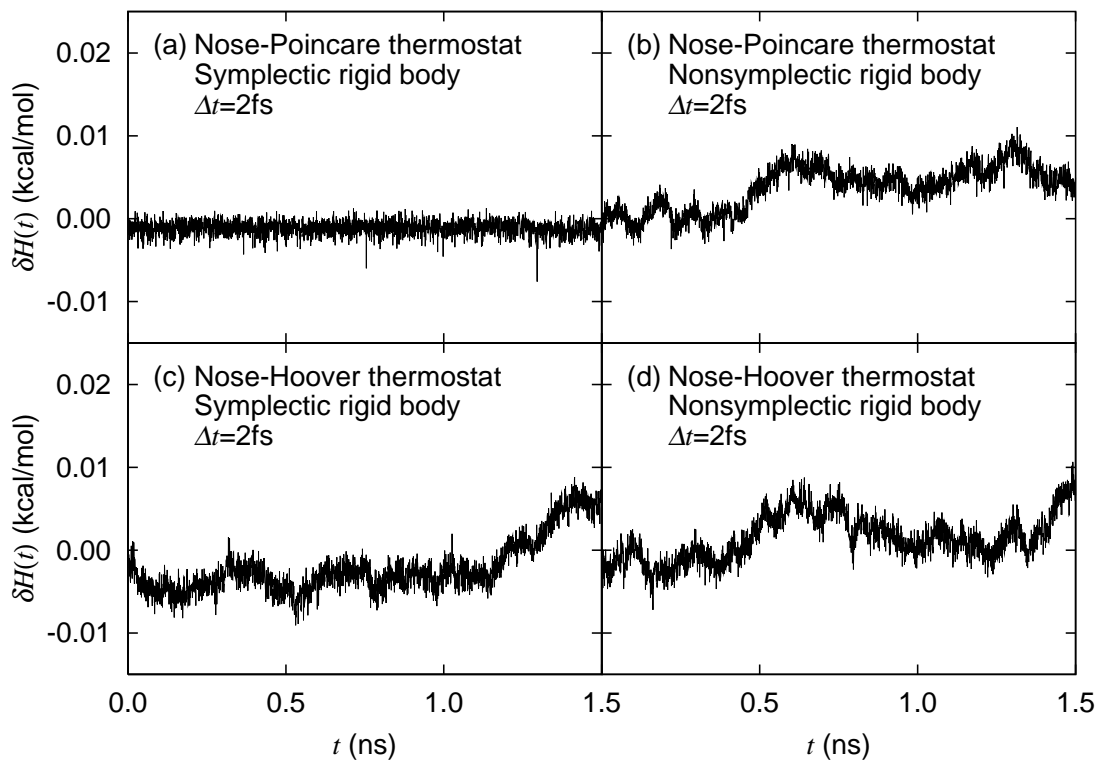}
\caption{
The time series of the difference of Hamiltonian 
from its initial value $\delta H(t)$. 
The time step was set to $\Delta t$ = 2~fs. 
(a) Nos\'e-Poincar\'e thermostat and symplectic rigid-body MD, 
(b) Nos\'e-Poincar\'e thermostat and nonsymplectic rigid-body MD, 
(c) Nos\'e-Hoover thermostat and symplectic rigid-body MD, and 
(d) Nos\'e-Hoover thermostat and nonsymplectic rigid-body MD. 
}
\label{1:fig}
\end{figure}

\begin{figure}
\includegraphics[angle=-90,width=15cm,keepaspectratio]{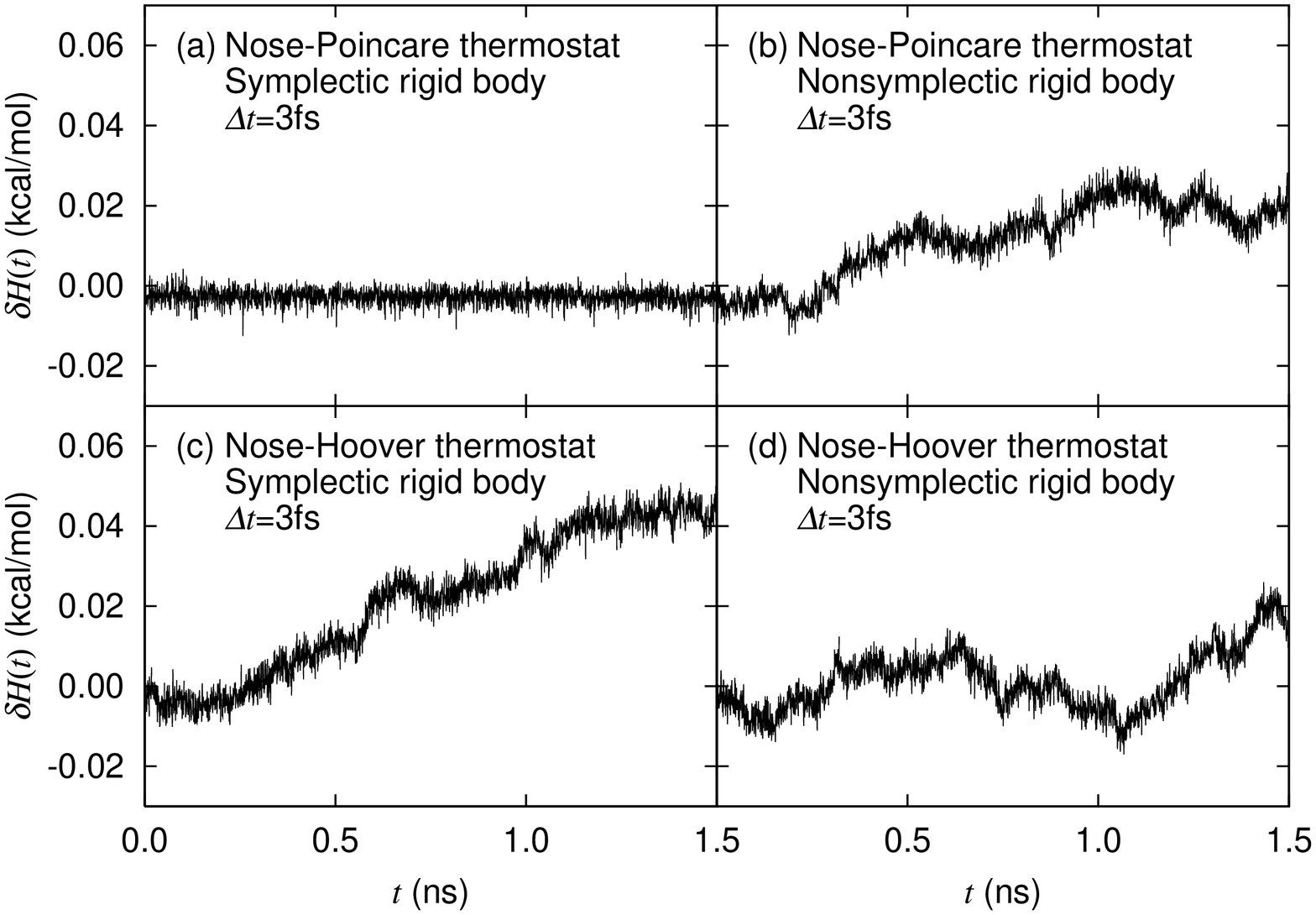}
\caption{
The time series of $\delta H(t)$. 
The time step was set to $\Delta t$ = 3~fs. 
See the caption of Fig.~1 for further details. 
}
\label{2:fig}
\end{figure}

\begin{figure}
\includegraphics[angle=-90,width=15cm,keepaspectratio]{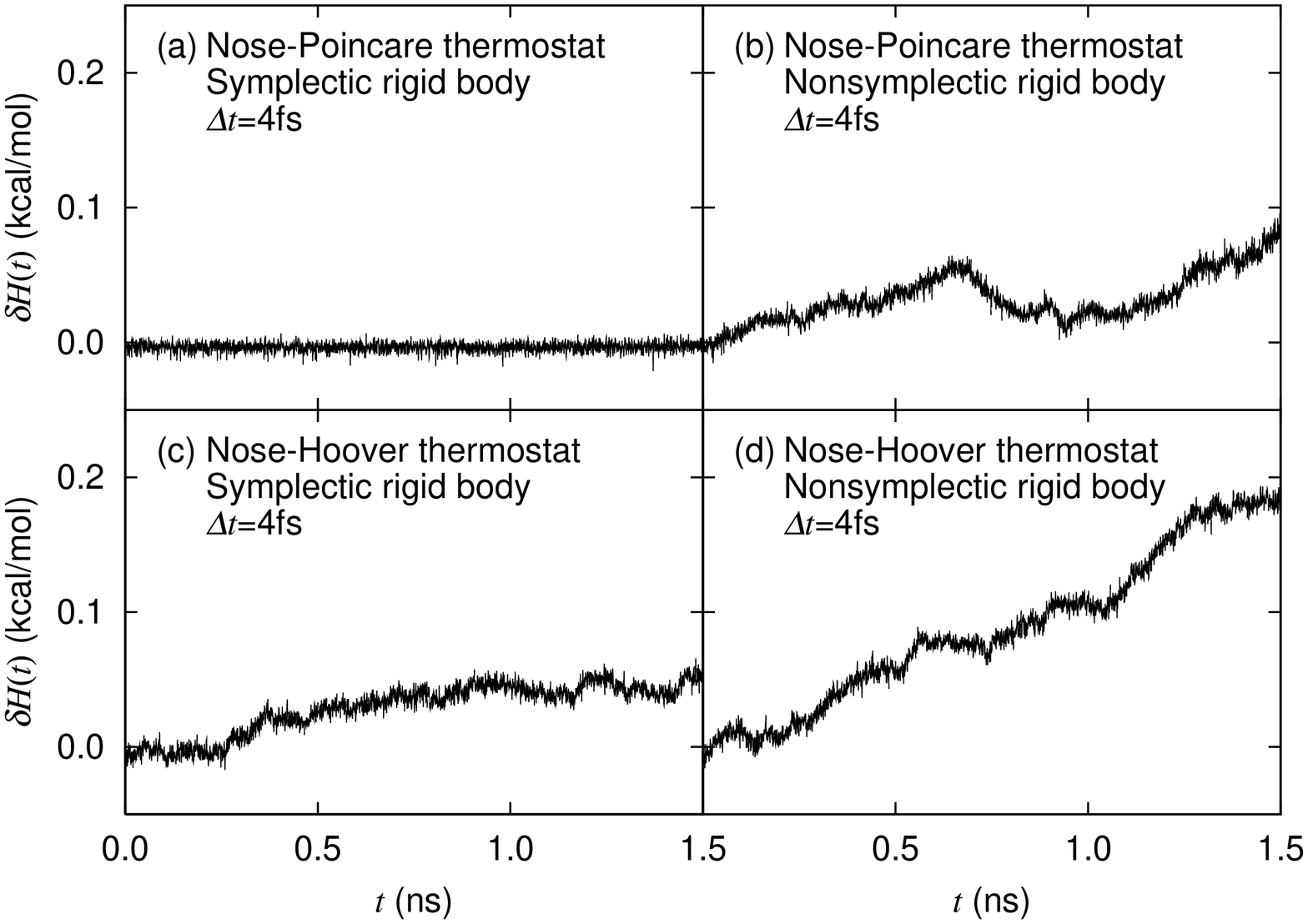}
\caption{
The time series of $\delta H(t)$. 
The time step was set to $\Delta t$ = 4~fs. 
See the caption of Fig.~1 for further details. 
}
\label{3:fig}
\end{figure}

\begin{figure}
\includegraphics[angle=-90,width=15cm,keepaspectratio]{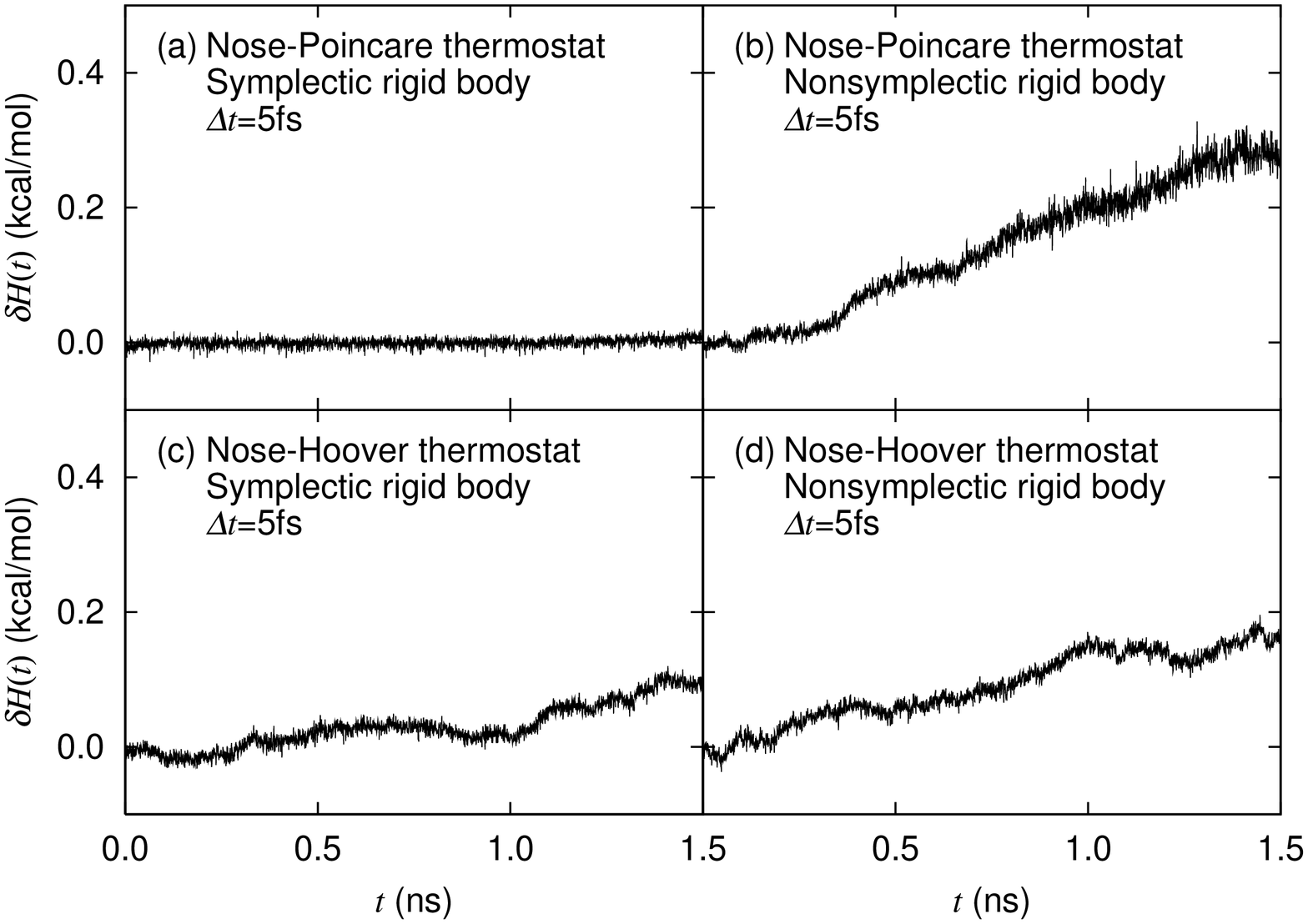}
\caption{
The time series of $\delta H(t)$. 
The time step was set to $\Delta t$ = 5~fs. 
See the caption of Fig.~1 for further details. 
}
\label{4:fig}
\end{figure}

We observed the deviations of the Nos\'e Hamiltonian from 
its initial values: 
\begin{equation}
\delta H (t) = 
  \sum_{i=1}^N \frac{\bs{p}'^2_i}{2m_is^2} + 
  \sum_{i=1}^N \frac{1}{8s^2} \bs{\pi}'^{\rm T}_i {\stackrel{\leftrightarrow}{\bs S}}(\bs{q}_i) 
               {\stackrel{\leftrightarrow}{\bs D}}_i {\stackrel{\leftrightarrow}{\bs S}}^{\rm T}(\bs{q}_i) \bs{\pi}'_i 
+ E(\bs{r}^{\{N\}}, \bs{q}^{\{N\}}) 
+ \frac{P_s^2}{2Q} + g k_{\rm B}T_0 \log s - H_0~. 
\end{equation}
Figures~\ref{1:fig}, \ref{2:fig}, \ref{3:fig}, and \ref{4:fig} show 
$\delta H (t)$ for $\Delta t$ = 2~fs, 3~fs, 4~fs, and 5~fs, respectively. 
The gradient of the linear fitting 
for each $\delta H (t)$ is shown in Table~\ref{1:table}.

In every nonsymplectic MD algorithm, 
Hamiltonian deviates from its initial value 
as time passes even for $\Delta t$ = 2~fs 
as shown in Figs.~\ref{1:fig}(b)-(d). 
This deviation increases as
the time step increases from $\Delta t$ = 2~fs to 5~fs. 
Note that the energy scale in the ordinate increases 
as $\Delta t$ increases. 

On the other hand, 
the Nos\'e-Poincar\'e thermostat and symplectic rigid-body MD algorithm 
guarantees the existence of a conserved quantity 
which is close to the Hamiltonian. 
Because of this conserved quantity, 
the Hamiltonian was conserved well 
for time steps of $\Delta t$ = 2~fs, 3~fs, and 4~fs 
as shown in Figs.~\ref{1:fig}(a)-\ref{3:fig}(a). 
The Hamiltonian starts to deviate slightly by 
$d \delta H (t)/ d t = 3.7 \times 10^{-3}$ kcal/mol/ns 
in the case of $\Delta t$ = 5~fs as shown in Table~\ref{1:table}. 
However, it is only two or five percent of $\delta H$ for 
the other nonsymplectic MD simulations (Table~\ref{1:table}). 
This fact means that employing the combination of 
the Nos\'e-Poincar\'e thermostat 
and the symplectic rigid-body algorithm, 
one can take a time step of as much as 4~fs. 
This time step is longer than typical values of 0.5~fs to 2~fs 
which are used by the conventional nonsymplectic algorithms.

%
  \section{Conclusions} \label{conc:sec}
We have proposed an explicit symplectic MD algorithm 
for rigid-body molecules in the canonical ensemble. 
This algorithm is based on 
the Nos\'e-Poincar\'e thermostat \cite{bond99,nose01} and 
the symplectic rigid-body algorithm \cite{mill02}. 
We also have presented an explicit symplectic MD algorithm 
for rigid-body molecules in the isothermal-isobaric ensembles 
by combining the Andersen barostat \cite{and80} 
with the symplectic algorithm in the canonical ensemble. 
As a modification of the isothermal-isobaric algorithm, 
we further presented the symplectic integrator 
in the constant normal pressure and 
lateral surface area ensemble 
and a symplectic algorithm combined with the Parrinello-Rahman algorithm. 
Employing the symplectic MD algorithm, there is a conserved quantity 
which is close to the Hamiltonian. 
Therefore, we can perform a MD simulation more stably 
than by conventional nonsymplectic algorithms.

In order to establish this fact numerically, 
we have applied this algorithm to a TIP3P pure water system at 300~K and 
compared the time evolution of the Hamiltonian 
with those by the nonsymplectic algorithms. 
These nonsymplectic algorithms are 
based on 
the Nos\'e-Poincar\'e thermostat \cite{bond99,nose01} and 
the nonsymplectic rigid-body algorithm \cite{matu99}, 
based on 
the Nos\'e-Hoover thermostat \cite{martyna96} and 
the symplectic rigid-body algorithm \cite{mill02}, and 
based on 
the Nos\'e-Hoover thermostat \cite{martyna96} and 
the nonsymplectic rigid-body algorithm \cite{matu99}. 
In these nonsymplectic algorithms, 
the Hamiltonian deviates gradually from its initial value 
in all cases of the time steps $\Delta t$ = 2~fs, 3~fs, 4~fs, and 5~fs. 
On the other hand, 
the Hamiltonian was conserved well even for a time step of 4~fs 
in our symplectic algorithm. 

The rigid-body model for molecules can be employed 
not only for a water system but also for a biomolecular system. 
For example, a partial rigid-body model \cite{ike04} is often used 
for a part of a peptide and a protein, 
in particular, for a hydrogen-including part 
to alleviate a fast motion of the hydrogen atom. 
Our algorithms will thus be of great use for 
MD simulations of an aqueous solution and a biomolecular system 
at a constant temperature and/or pressure.

%
\section*{ACKNOWLEDGEMENTS}
This work was supported, in part,
by the Grants-in-Aid for the Next Generation 
Super Computing Project, Nanoscience Program and
for Scientific Research in Priority Areas, ``Water and Biomolecules'',
from the Ministry of Education, Culture, Sports, 
Science and Technology, Japan.

%
%
%

%
%
%

\vspace{3cm}
    
\begin{table}[h]
\caption{
Drift of the Hamiltonian per nanosecond $d \delta H/dt$ (kcal/mol/ns).}
\label{1:table}
\label{ham:table}
\begin{center}
\begin{tabular}{ccccc} \hline   
  $\Delta t$ & 2~fs & 3~fs & 4~fs & 5~fs \\ \hline
 Nos\'e-Poincar\'e and symplectic rigid-body MD 
 &   $-1.4 \times 10^{-4}$ &   $-3.0 \times 10^{-4}$ & 
  \ \ $2.0 \times 10^{-4}$ &\ \ $3.7 \times 10^{-3}$ \\
 Nos\'e-Poincar\'e and nonsymplectic rigid-body MD 
 &\ \ $4.4 \times 10^{-3}$ &\ \ $1.9 \times 10^{-2}$ & 
  \ \ $2.8 \times 10^{-2}$ &\ \ $2.2 \times 10^{-1}$ \\
 Nos\'e-Hoover and symplectic rigid-body MD 
 &\ \ $5.3 \times 10^{-3}$ &\ \ $3.9 \times 10^{-2}$ & 
  \ \ $3.8 \times 10^{-2}$ &\ \ $6.9 \times 10^{-2}$ \\
 Nos\'e-Hoover and nonsymplectic rigid-body MD 
 &\ \ $2.9 \times 10^{-3}$ &\ \ $7.3 \times 10^{-3}$ & 
  \ \ $1.3 \times 10^{-1}$ &\ \ $1.2 \times 10^{-1}$ \\
 \hline
\end{tabular}
\end{center}
\end{table}
%
%
%
%
%
%
%
%
%
%
%
%
%
\end{document}